\documentclass[twocolumn,english,twocolumn,nobibnotes,nofootinbib]{revtex4-1}
\usepackage[T1]{fontenc}
\usepackage[latin9]{inputenc}
\usepackage{amsmath}
\usepackage{amssymb}
\usepackage{graphicx}
\usepackage{esint}

\makeatletter
\@ifundefined{textcolor}{}
{%
 \definecolor{BLACK}{gray}{0}
 \definecolor{WHITE}{gray}{1}
 \definecolor{RED}{rgb}{1,0,0}
 \definecolor{GREEN}{rgb}{0,1,0}
 \definecolor{BLUE}{rgb}{0,0,1}
 \definecolor{CYAN}{cmyk}{1,0,0,0}
 \definecolor{MAGENTA}{cmyk}{0,1,0,0}
 \definecolor{YELLOW}{cmyk}{0,0,1,0}
}

\@ifundefined{definecolor}
 {\usepackage{color}}{}

\usepackage{float}\usepackage{babel}
\usepackage{braket}

\makeatother

\usepackage{babel}
\begin{document}

\author{K. Kolasi\'{n}ski, }

\affiliation{AGH University of Science and Technology, Faculty of Physics and
Applied Computer Science,\\
 al. Mickiewicza 30, 30-059 Kraków, Poland}

\author{B. Szafran}

\affiliation{AGH University of Science and Technology, Faculty of Physics and
Applied Computer Science,\\
 al. Mickiewicza 30, 30-059 Kraków, Poland}

\author{B. Brun }

\affiliation{Université Grenoble Alpes, F-38000 Grenoble, France \\
 CNRS, Institut NEEL, F-38042 Grenoble, France }

\author{H. Sellier}

\affiliation{Université Grenoble Alpes, F-38000 Grenoble, France \\
 CNRS, Institut NEEL, F-38042 Grenoble, France }

\title{Interference features in scanning gate conductance maps of quantum
point contacts with disorder}
\begin{abstract}
We consider quantum point contacts (QPCs) defined within disordered two-dimensional
electron gases as studied by scanning gate microscopy. We evaluate
the conductance maps in the Landauer approach and wave function picture
of electron transport for samples with both low and high electron
mobility at finite temperatures. We discuss the spatial distribution
of the impurities in the context of the branched electron flow. We
reproduce the surprising temperature stability of the experimental
interference fringes far from the QPC. Next, we discuss -- previously
undescribed -- funnel-shaped features that accompany splitting of
the branches visible in previous experiments. Finally, we study elliptical
interference fringes formed by an interplay of scattering by the point-like
impurities and by the scanning probe. We discuss the details of the
elliptical features as functions of the tip voltage and the temperature,
showing that the first interference fringe is very robust against
the thermal widening of the Fermi level. We present a simple analytical
model that allows for extraction of the impurity positions and the
electron gas depletion radius induced by the negatively charged tip
of the atomic force microscope, and apply this model on experimental
scanning gate images showing such elliptical fringes.
\end{abstract}
\maketitle

\section{Introduction}

Scanning gate microscopy (SGM) is an experimental technique which
probes transport properties of systems based on a two-dimensional
electron gas (2DEG) using the charged probe of an atomic force microscope
(AFM) \cite{Eriksson1996,Sellier2011,Ferry2011}. A negatively charged
AFM tip induces a finite size depletion in the 2DEG, which acts as
a movable scatterer of size and location controlled by the voltage
applied to the SGM tip and its position above the sample \cite{Brun2014}.
The SGM technique was used first to investigate the electron transport
in quantum point contacts (QPCs) \cite{Wees1988}. The SGM conductance
maps recorded as a function of the tip position in the vicinity of
a QPC contain two characteristic features (i) interference fringes
with oscillation period equal to half the Fermi wavelength $\lambda_{\mathrm{F}}$
\cite{Topinka2000,Topinka2001,Kozikov2015,Jura2007,Jura2009,Kumar2013}
and (ii) semiclassical branched flow of electron trajectories \cite{Heller2003,Liu2013,Liu2015,Jura2007}.
The fringes (i) arise from the coherent interference between the electron
waves incident from the QPC and backscattered by the SGM tip \cite{Kolasinski2014Slit,Jura2007}.
The branched flow (ii) stems from the smooth potential disorder in
the high mobility semiconductor structures \cite{Jura2007}. For low-mobility
samples the hard-impurity scattering is dominant and leads to coherent
fringes which are surprisingly thermally stable, with the interference
pattern visible at a distance from the QPC which largely exceeds the
thermal length $\lambda_{\mathrm{th}}$ \cite{Topinka2000,Topinka2001,Jura2007}.
This surprising behavior is explained \cite{Topinka2000,Topinka2001,Jura2007}
by coherent scattering involving the tip and nearby impurities spaced
by a distance below $\lambda_{\mathrm{th}}$. 

In this paper we consider numerical simulations of a coherent branched
flow of electrons spreading from a QPC into the 2DEG. We study the
effect of smooth and hard impurities on the transport for both low
and high density of scatterers, i.e. for high and low mobility samples,
respectively. For low mobility samples most of the features visible
in the experimental SGM images can be explained in terms of a 1D model
of the branch, including (i) thermally {persistent} fringes visible
at $T\simeq4$K, (ii) reappearance of fringes in some part of SGM
images far from the QPC, (iii) perpendicular alignment of the fringes
to the branch direction, (iv) frequency of the fringes near the impurities
that changes with $T$. We discuss splittings of the branches at some
defects and a funnel shaped features that accompany the splitting.

We also consider high mobility samples and indicate by both experiment
and theory distinct signatures of a few hard scatterers present within
the system that produce pronounced elliptical features in the SGM
conductance maps. These elliptical features -- never previously described
-- result from interference involving both the scatterer and the tip
and remain stable up to at least $T\simeq4$ K. We provide a simple
model to describe these nearly elliptical contours which allows one
to indicate the position of the scatterer within the sample, and the
size of the area depleted by the tip.



\section{Model }

We consider a 2DEG system with a local constriction formed by the
QPC {[}Fig. 1{]}. The electrons are fed from the input lead at left
of the QPC. Behind the QPC the electrons propagate freely, with open
boundary conditions denoted by arrows at the blue edge of Fig. 1(a).
We consider the scattering of the Fermi level electrons solving the
effective-mass Schrödinger equation (atomic units are used) 
\begin{equation}
-\frac{1}{2m_{\mathrm{eff}}}\nabla^{2}\Psi+eV_{\mathrm{tot}}\Psi=E_{\mathrm{F}}\Psi,\label{eq:Ham}
\end{equation}
 where $\Psi\equiv\Psi\left(x,y\right)$ is the two-dimensional scattering
wave function with density $\rho=\left|\Psi\right|^{2}$, and $m_{\mathrm{eff}}=0.067$
is GaAs electron effective mass. In Eq. (1), $V_{\mathrm{tot}}=V_{\mathrm{QPC}}+V_{\mathrm{tip}}+V_{\mathrm{dis}}$
contains contributions of all possible sources of electrostatic potentials
considered in this paper. We assume $E_{F}=15$ meV which corresponds
to 2DEG density of $4.2\times10^{11}$ /cm$^{2}$. 
 $V_{\mathrm{QPC}}$ is the QPC electrostatic potential modeled with
Davies formula \cite{Davies1995} for a finite rectangular gate (green
rectangles on Fig. 1(a)) 
\begin{align*}
V_{\mathrm{QPC}}/V_{\mathrm{g}} & =g(x-L,y-B)+g(x-L,T-y)\\
 & +g(R-x,y-B)+g(R-x,T-y)
\end{align*}
where $g(u,v)=\frac{1}{2\pi}\arctan\left(\frac{uv}{dP}\right);\,\, P=\sqrt{v^{2}+u^{2}+d^{2}}$,
with $L$, $R$, $B$ and $T$ being the left, right, bottom and top
position of the gate edges, (see Fig. 1(a)). We choose the distance
between 2DEG and gates to be $d=50$nm. In the above formula $V_{\mathrm{g}}$
is the gate potential. For the applied parameters, the Fermi energy
$E_{F}=15$ meV corresponds to the first conductance plateau of the
QPC. $V_{\mathrm{tip}}$ is the electrostatic potential of the charged
tip, for which we use the Lorentzian approximation, 
\begin{equation}
V_{\mathrm{tip}}=\frac{d_{\mathrm{tip}}^{2}V_{\mathrm{t}}}{\left(x-x_{\mathrm{tip}}\right)^{2}+\left(y-y_{\mathrm{tip}}\right)^{2}+d_{\mathrm{tip}}^{2}},\label{eq:lorentz}
\end{equation}
The Lorentzian form of the tip potential arises due to screening by
the electron gas inside the heterostructure \cite{kolasinskiDFT2013,szafranDFT2011,Steinacher2015}.
The width of the tip is of order of the tip - 2DEG distance and fixed
at $d_{\mathrm{tip}}=80$nm. The maximum potential change $V_{\mathrm{t}}$
induced by the tip in the 2DEG is taken to be 30meV (except otherwise
stated) corresponding to a depletion area of radius $d_{\mathrm{tip}}$.
This simple form of tip potential corresponds to the case of linear
screening by the 2DEG electrons \cite{kolasinskiDFT2013,szafranDFT2011},
while the more complicated case with 2DEG depletion (i.e. when $V_{\mathrm{t}}>E_{\mathrm{F}}$)
would require self consistent numerical calculations. Finally, the
last contribution to the potential, $V_{\mathrm{dis}}$ arises from
the disorder in the donor layer and it is assumed to be a superposition
of uniformly distributed Gaussian functions 
\[
V_{\mathrm{dis}}\left(x,y\right)=\sum_{i=1}^{N_{\mathrm{dis}}}\alpha_{\mathrm{i}}e^{-\left|r-r_{\mathrm{i}}\right|^{2}/2\sigma^{2}},
\]
 where $N_{\mathrm{dis}}$ is the number of impurities, $\left|\alpha_{\mathrm{i}}\right|<V_{\mathrm{max}}$
is the potential amplitude generated from a uniform random distribution
, $r_{\mathrm{i}}$ is the center of the i-th Gaussian randomly distributed
in the device and $\sigma$ is the width of the Gaussians. We use
$\sigma=12$nm and $V_{\mathrm{max}}=0.3\times E_{\mathrm{F}}$ and
$\alpha_{\mathrm{i}}>0$ for \emph{hard} impurities, and $\sigma=30$nm
and $V_{\mathrm{max}}=0.05\times E_{\mathrm{F}}$, for smooth impurities
(see Fig. 1(b,c)). 

We use the finite difference discretization of Eq. \eqref{eq:Ham}
and wave function matching (WFM) -- described in the Appendix -- in
order to include the effect of the leads into the Hamiltonian and
calculate the scattering amplitudes \cite{Zwierzycki2008,AndoWFM1991,Khomaykov2005}.
The conductance of the system is then calculated from the Landauer
formula 
\begin{equation}
G(E_{\mathrm{F}},T)=G_{\mathrm{0}}\int_{-\infty}^{+\infty}dEM(E)\left(-\frac{\partial f(E;E_{\mathrm{F}},T)}{\partial E}\right),\label{eq:G}
\end{equation}
 with $f(E;E_{\mathrm{F}},T)=1/\left(\exp\left(\left(E-E_{\mathrm{F}}\right)/k_{\mathrm{B}}T\right)+1\right)$
being the Fermi-Dirac distribution, $M(E)=\sum_{i=1}^{M}T_{i}(E)$
is the total transmission summed over all incoming modes in the input
lead and $G_{\mathrm{0}}=2e^{2}/h$ is the conductance quantum.

\begin{figure}
\begin{centering}
\includegraphics[width=1\columnwidth]{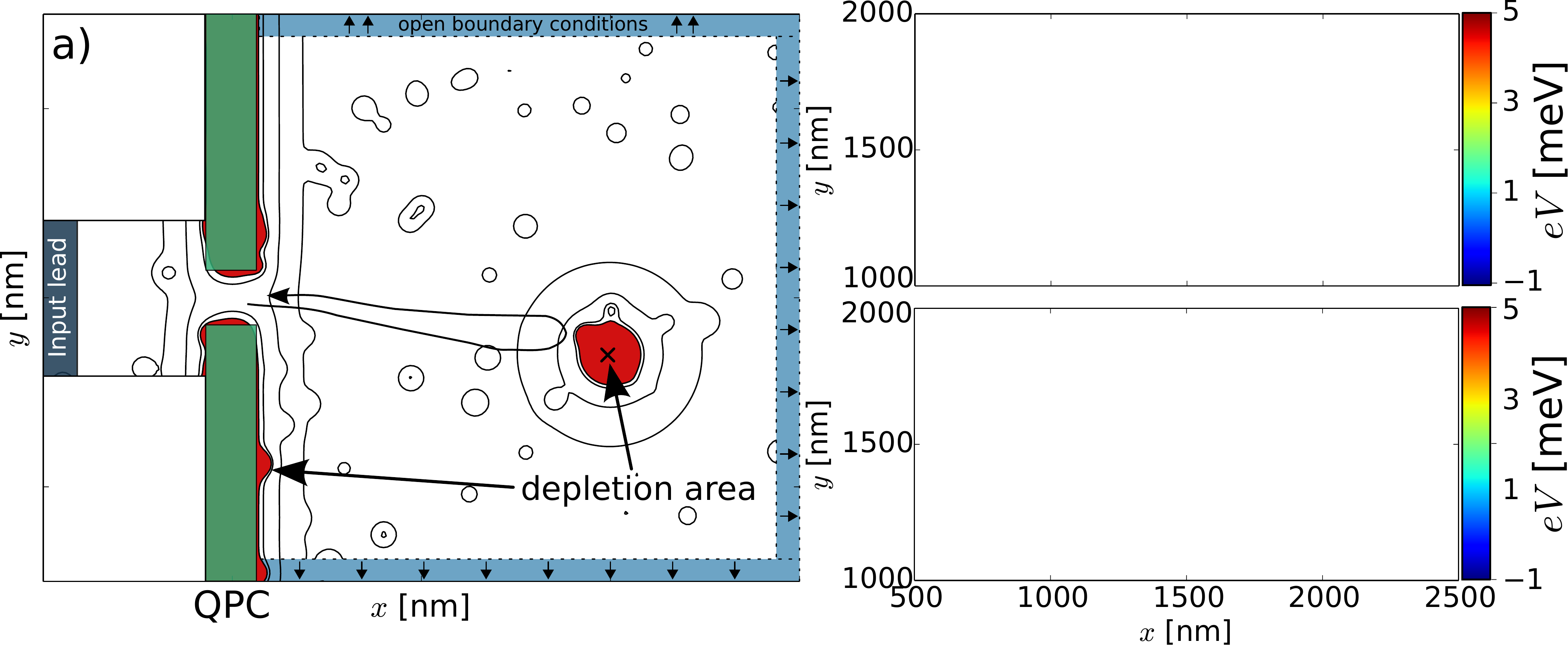} 
\par\end{centering}

\caption{a) The sketch of the system. The loop marked by the black arrow schematically shows the scattering
process leading to the $\lambda_{\mathrm{F}}/2$ fringes in SGM images.
The brown map shows the calculated scattering electron density ploted
as $\sqrt{\rho}$. Disordered potential landscape in the sample near
QPC for case of a b) \emph{smooth} impurities c) \emph{hard} impurities.
White solid lines in (b) and (c) show the isolines of the potential
energy for $eV_{\mathrm{tot}}=E_{\mathrm{F}}$. The QPC is tuned to
the first plateau.}
\end{figure}

\section{Results}

\subsection{Effect of disorder on the SGM maps}

\begin{figure}
\begin{centering}
\includegraphics[width=1\columnwidth]{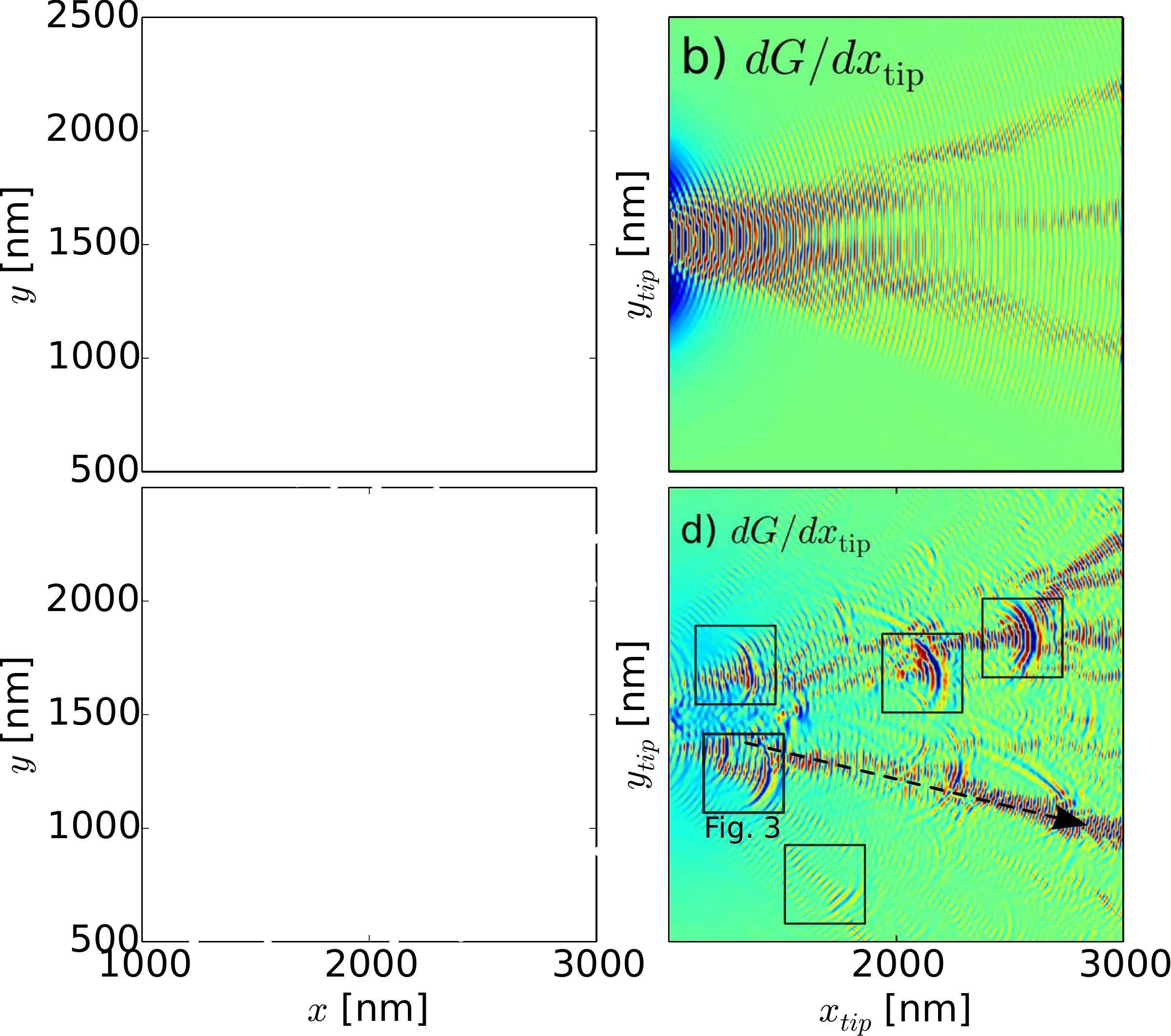} 
\par\end{centering}

\caption{\label{fig:dis}a) Square root of scattering electron density $\sqrt{\rho}$
for the case of smooth impurity background potential. b) Resulting
$dG/dx$ SGM image. c-d) Same as a) and b) but for hard impurities.
Squares in c) and d) denote the funnel like fringe pattern discussed
in the text. In (c,d) we considered $N_{d}=500$ impurities within
the entire computational box. In (a,b) we kept all the impurities
of (c-d) and inserted another 500 placed at random positions.}
\end{figure}

In Figure \ref{fig:dis}(a) we show the scattering electron density
for the smooth disorder at $T$=0K with $dG/dx$ depicted in Fig. \ref{fig:dis}(b).
A branched flow is formed far from the QPC with well visible $\lambda_{\mathrm{F}}/2$
fringes \cite{Jura2007}. Near the QPC characteristic circular fringes
\cite{Kolasinski2014Slit} appear due to the standing wave between
the QPC and the tip {[}see the backscattered trajectory in Fig. 1(a){]}.
Smooth defects lead to small-angle scattering and the branched flow
remains straight over large distances. This kind of flow is found
in the high mobility samples \cite{Jura2007}.

Figure \ref{fig:dis}(c) shows the scattering electron density for
the case of hard impurities. The potential centers (white dots) are
superimposed on the electron density image in order to show the relation
between location of branches and impurity distribution within the
sample. From this image one notices that the two main branches are
formed along the lines with a lower impurity density (one of those
branches is denoted by the black arrow in Fig. \ref{fig:dis}(d)).
Not every impurity splits  the electron flow in
branches and the current passes across some of them. 

\begin{figure}
\begin{centering}
\includegraphics[width=1\columnwidth]{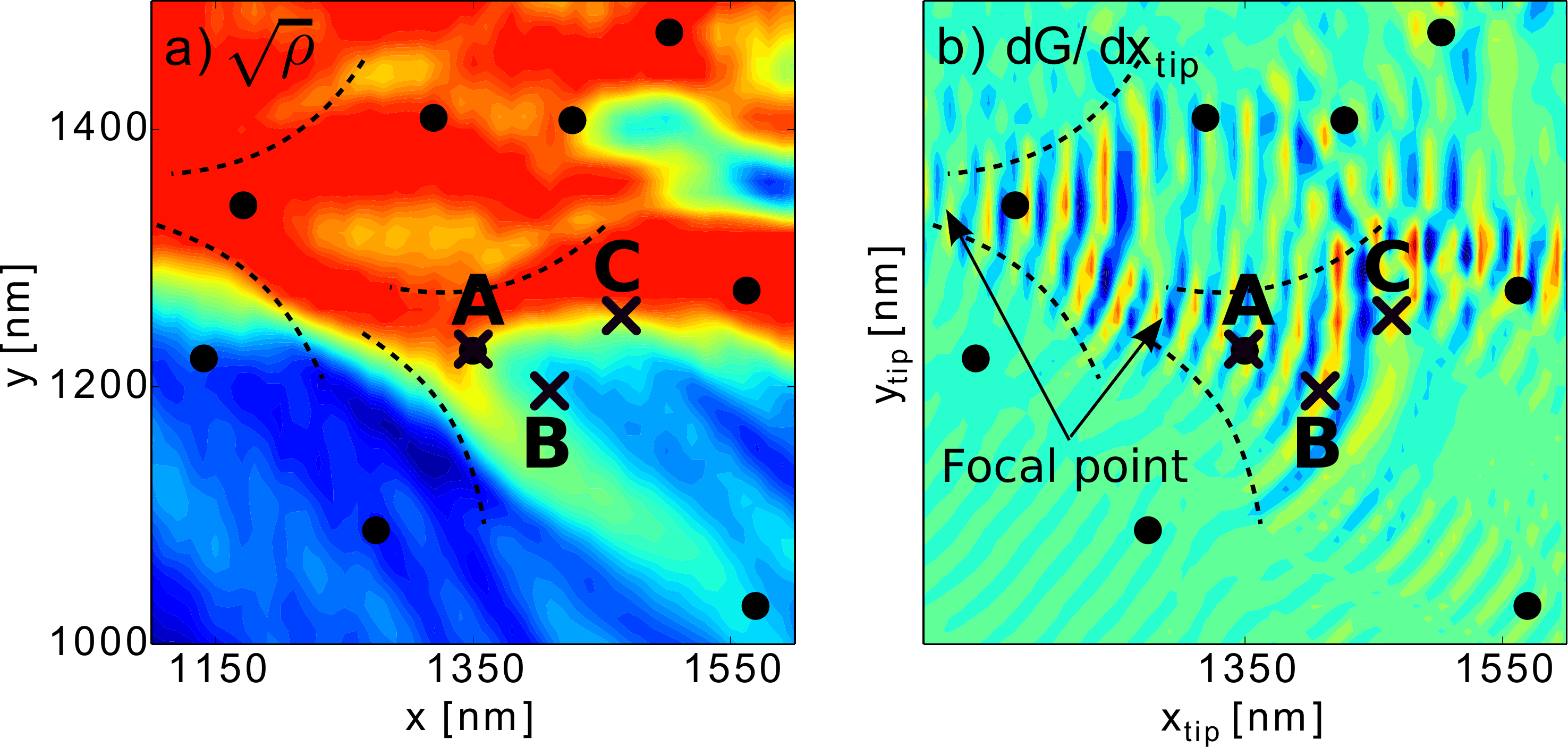} 
\par\end{centering}

\caption{\label{fig:zoom}(a-b) Zoom of area denoted by \emph{``}Fig. 3\emph{''}
square in Fig. \ref{fig:dis}(d). Points A, B and C denote the possible
tip positions of the scattering scenario schematically presented in
Figs. \ref{fig:proc}(b), (c) and (d). }
\end{figure}

\begin{figure}
\begin{centering}
\includegraphics[width=1\columnwidth]{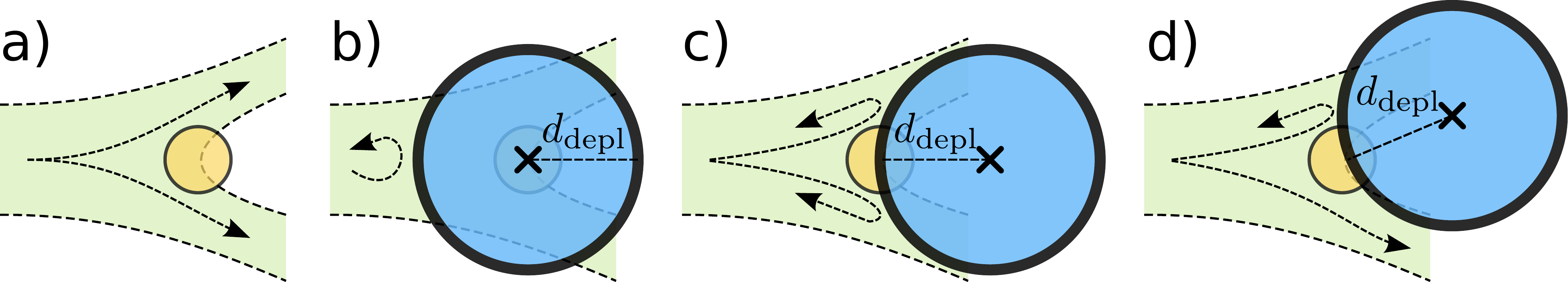} 
\par\end{centering}

\caption{\label{fig:proc}Sketch of the scattering process leading to the funnel-shaped
fringes in the SGM images. a) Branched flow around impurity without
SGM tip. b) When tip is above impurity, c) at the widest branch point
when the tip still closes both paths and d) when the tip deletion
area leaves the funnel-shaped branch around the impurity. The $d_{\mathrm{delp}}$
is the radius of the area depleted by the SGM tip.}
\end{figure}

Aside from the two dominating branches in Fig. \ref{fig:dis}(c) one
can see a number of characteristic funnel shaped fringe patterns denoted
by the squares. Those patterns accompany the splitting of the electron
density in two branches by a hard impurity in the branch. 
This process is sche\-matically presented in Fig. \ref{fig:proc}(a)
and can be also noted in the Fig. \ref{fig:zoom}(a). Due to the finite
size of the obstacle, the electron has to flow around, which leads
to the funnel-shaped local widening of the branch near the impurity.
In the presence of the SGM tip the electron waves can be backscattered
within the funnel area (see Fig. \ref{fig:proc}(b) and (c)), which
results in the characteristic circular fringes visible in Fig.\ref{fig:zoom}(b).
At some point when the tip depletion area does not block both parts
of the split branches, backscattering is reduced and circular fringes
disappear from the SGM images. 
The funnel-shaped fringe patterns are also visible in the experimental
images e.g. see in Fig. 2(b) of Refs. \cite{Paradiso2010} and \cite{Topinka2001}.
Let us note that by analyzing the size of the circular fringes one
may roughly estimate the depletion radius $d_{\mathrm{depl}}$ induced
by the SGM tip, as the distance between the funnel focal point and
the last fringe in the funnel i.e. the distance between tip location
inducing the flow depicted in Figs. \ref{fig:proc}(b) and \ref{fig:proc}(d)
(or points A and C in Fig. 3). From Fig. \ref{fig:zoom}(b) we get
an approximated value of depletion radius $d_{\mathrm{depl}}\approx120$nm.
This value of the order of the one obtained from condition $E_{\mathrm{F}}=V_{\mathrm{tip}}$,
which is 
\begin{equation}
d_{\mathrm{depl}}=d_{\mathrm{tip}}\sqrt{\frac{V_{\mathrm{t}}}{E_{\mathrm{F}}}-1}=80\mbox{\ensuremath{\mathrm{nm}}}.\label{eq:ddepl}
\end{equation}


\subsection{Thermal stability of the fringes}

\begin{figure}
\begin{centering}
\includegraphics[width=1\columnwidth]{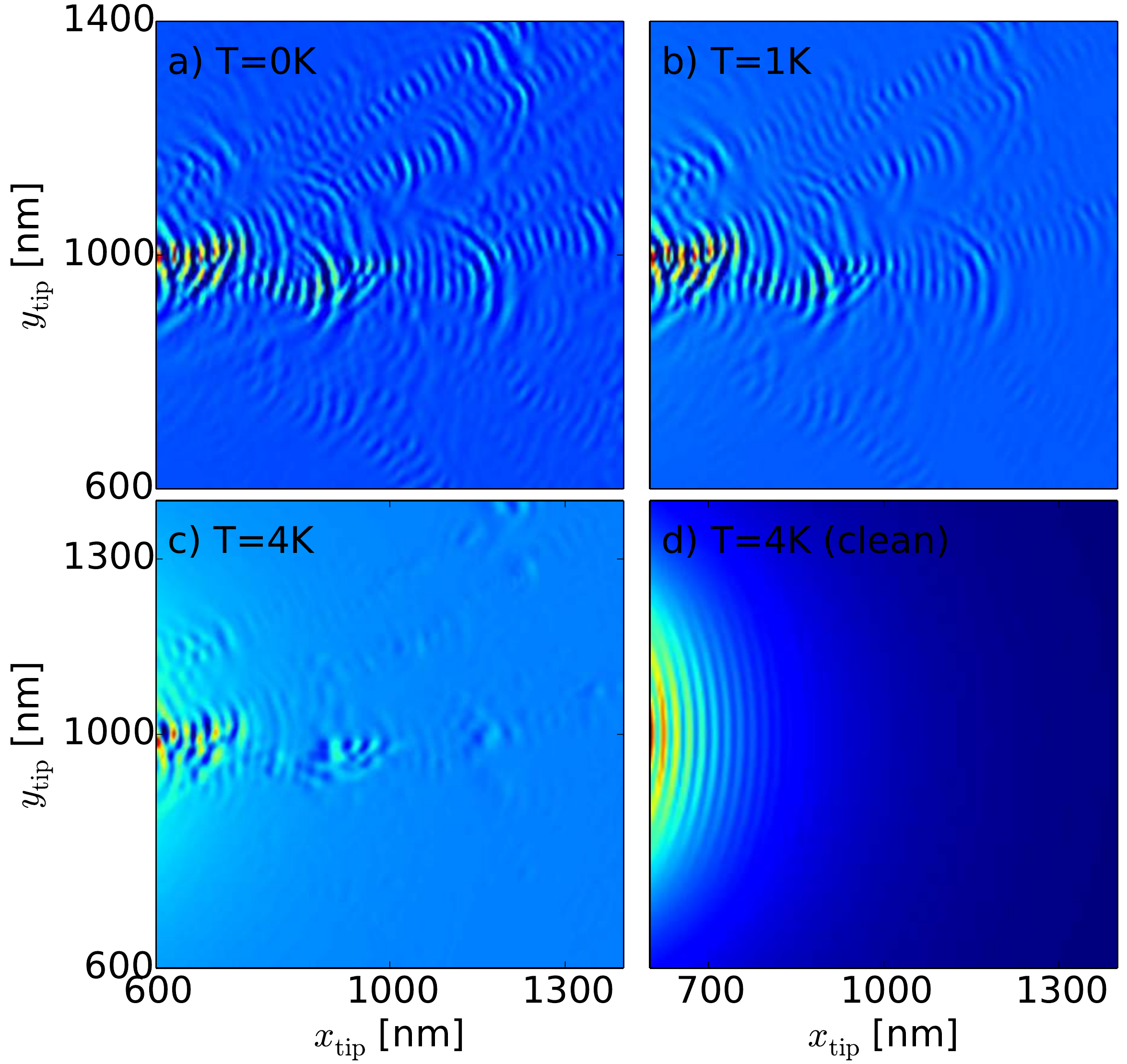} 
\par\end{centering}

\caption{\label{fig:branchTemp}(a-c) SGM images in presence of disorder obtained
at (a) T = 0 K, (b) T =1 K, and (c) T = 4 K. (d) Same as (c) without
disorder. The quantity $dG/dx$ is plotted in arbitrary units.}
\end{figure}

One of the most unexpected feature of the branched flow in the disordered
samples is the stability of the interference fringes against thermal
broadening which allows for observation of the fringes at several
microns from the QPC at $T=4$K, when the thermal length is only $l_{th}=\frac{2\pi\hbar^{2}}{m\lambda_{\mathrm{F}}k_{B}T}=400$
nm \cite{Topinka2000,Topinka2001,Jura2007} . 
In Figs. \ref{fig:branchTemp}(a-c) we show the simulated SGM $dG/dx$
maps for a system with hard impurities at $T=$0, 1, 4K. 

Comparing both Figs. \ref{fig:branchTemp}(c) and (d) one can see
that the persistence of the interference fringes at high temperatures
at large distances is directly caused by the disorder within the sample
\cite{Topinka2000,Topinka2001}. Additionally, a few other features
can be found in the SGM images: i) interference fringes are perpendicular
to the flow direction, 
ii) at $4K$ some fringes disappear for a short distance to reappear
further, iii) in general the fringe period is not uniform.

For the current flowing in branches the transport across the 2D system
can be reduced to the 1D scattering system provided that the current
leakage from the branch and the branch splittings are neglected. We
found that the observed features of the branches can be explained
within a model in which the electron branch is treated as a one dimensional
electron channel. 
The perpendicular orientation of the fringes inside the branch is
directly implied.

\begin{figure}
\begin{centering}
\includegraphics[width=1\columnwidth]{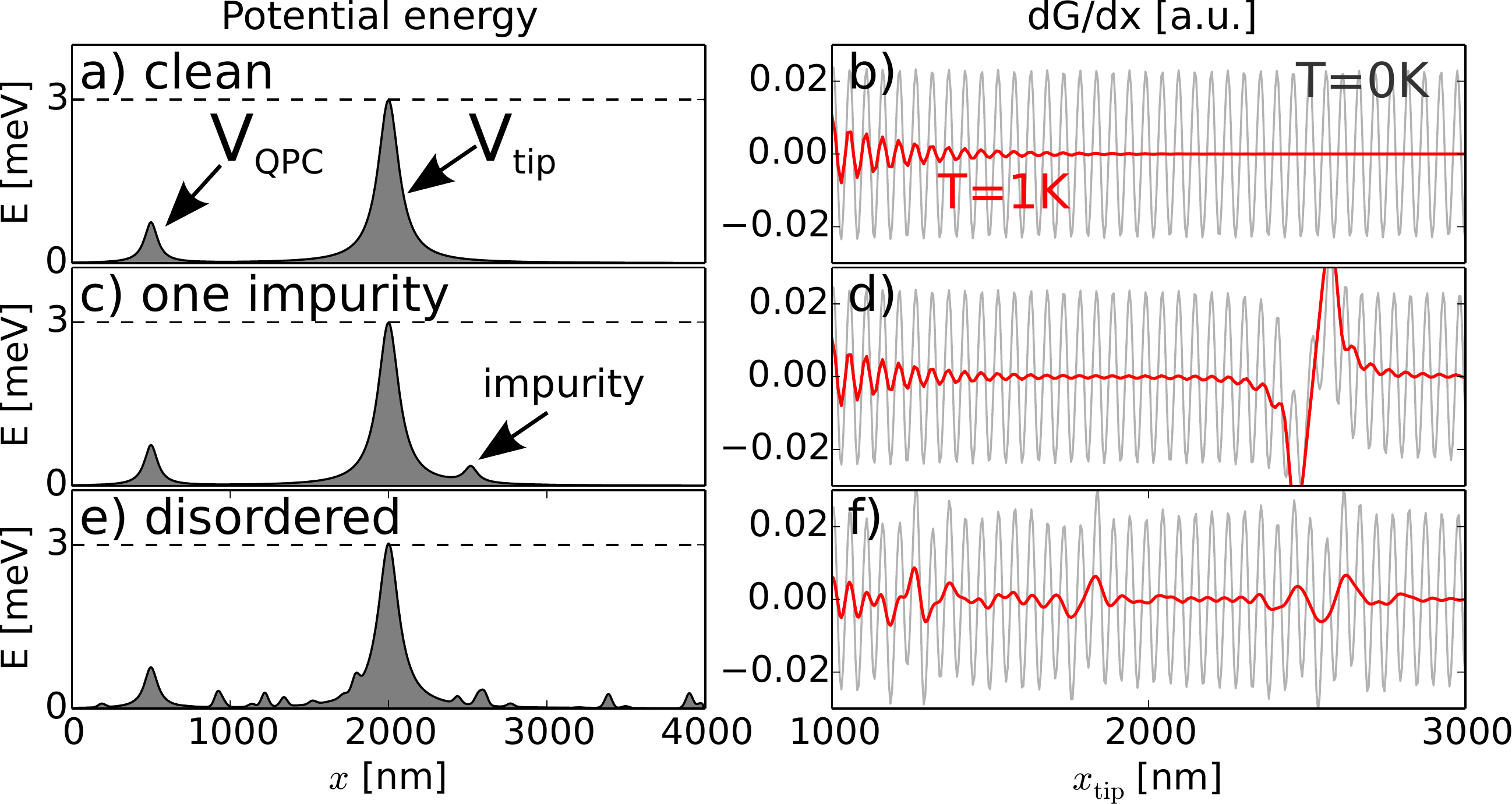} 
\par\end{centering}

\caption{\label{fig:1d}a) Potential energy landscape for clean 1D branch and
b) resulting $dG/dx$ conductance for $T$=0 and 1K. c-d) Same as a-b)
but for the case of one impurity inside the channel and e-f) for a
large number of impurities.}
\end{figure}

In Fig. \ref{fig:1d}(a) we show a 1D representation of a ``clean''
branch. An electron with kinetic energy $E_{\mathrm{F}}=3$meV is
incoming from the left reservoir and scatters on the QPC and SGM tip
potentials inside the channel. We set $V_{\mathrm{t}}=E_{\mathrm{F}}$.
The calculated SGM signal $dG/dx$ is presented in Fig. 6(b) for $T
$= 0 and 1 K. For $T=1$K the interference fringes disappear as functions
of tip position along the branch, which results from the finite width
of the transport window near the Fermi level. The simulation for a
single impurity within the channel {[}Fig. \ref{fig:1d}(c){]} shows
that the interference fringes reappear around the impurity {[}Fig.
\ref{fig:1d}(d){]}. This is possible when the distance between SGM
tip and the impurity becomes smaller than the thermal length \cite{Jura2007}.
The measured current is then sensitive to the interference which takes
place far from the QPC, thus the presence of fringes in the SGM images
at large distances is an evidence of nearby impurities. In Fig. \ref{fig:1d}(e-f)
we show that for a disordered channel the fringes remain visible at
large tip distances for $T=1$K, which results from the multiple scattering
between tip and nearby impurities. This effect is more dramatic in
case of 2D scattering, where for $T=$4K in Fig. \ref{fig:branchTemp}(c)
the amplitude of the fringes at some points is reduced almost to zero.
One may note that at some points temperature does change the period
of the fringes around the impurities in Fig. \ref{fig:1d}(f). The
non-uniformity of the fringe spacing at finite temperature was experimentally
observed for instance in Fig. 4 of Ref. \cite{Kozikov2015} and in
Fig. 7 of Ref. \cite{Kozikov2013}.

\subsection{A single hard scatterer in a high mobility sample \label{sub:hs}}

\begin{figure}
\begin{centering}
\includegraphics[width=1\columnwidth]{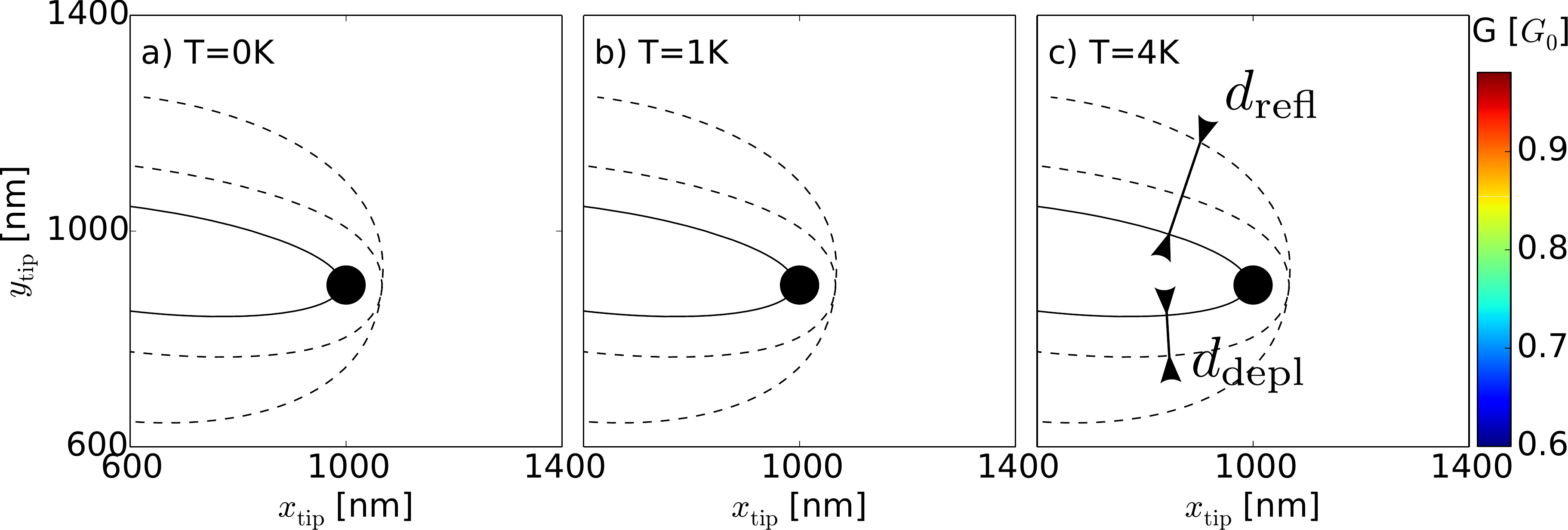} 
\par\end{centering}

\caption{\label{fig:imp}a) SGM image obtained for $T=$0 and one hard impurity
in the system with position denoted by circle. b) Same but for $T=1$K
and c) $T=4$K. The smallest ellipse is calculated from condition $r_{\mathrm{q-t-i}}-r_{\mathrm{q-i}}=\lambda_{\mathrm{F}}/2$.
The middle ellipse is the smallest one enlarged by $d_{\mathrm{depl}}$
along the normal direction. The largest contour is calculated from
the correction on the finite size of the tip depletion radius and
the scattering angle from Eq. \eqref{eq:depl}.}
\end{figure}

\begin{figure}
\begin{centering}
\includegraphics[width=1\columnwidth]{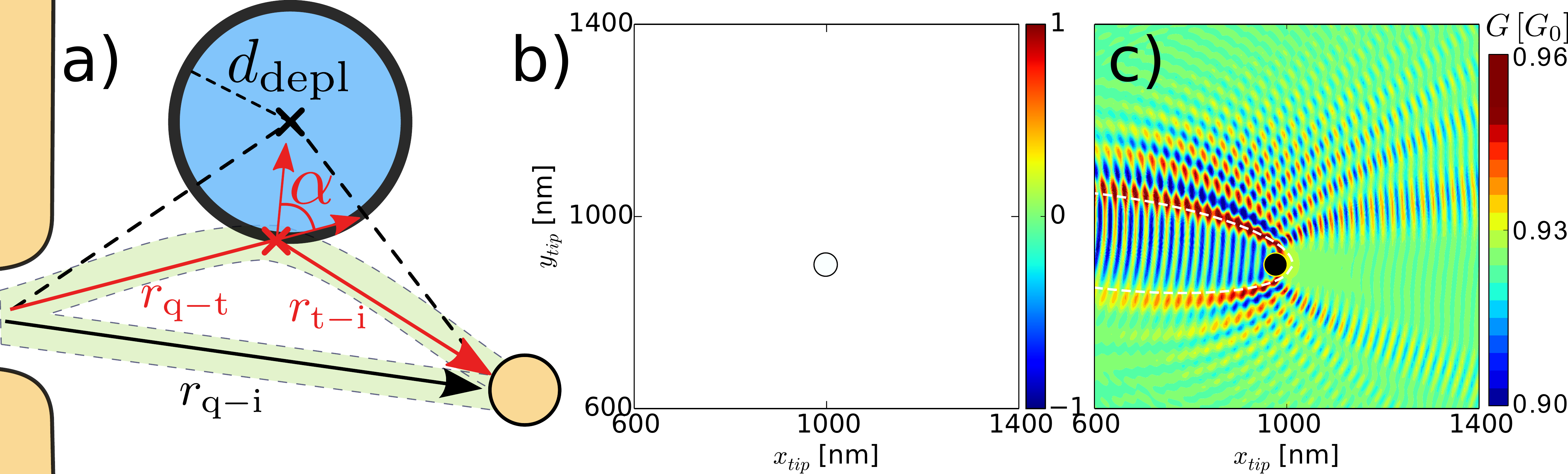} 
\par\end{centering}

\caption{\label{fig:elips}a) Two paths which lead to the elliptic fringe pattern
displayed in (b). The blue disk is the 2DEG depleted area beneath
the tip. The orange dot is the hard scatterer. The dashed lines show
an effective shift of the tip position that is due to the finite size
of the depletion area and $\alpha$ is the angle of incidence of the
electron wave to the depleted area. b) The elliptic fringes obtained
from Eq. \eqref{eq:ellips}. c) The SGM image obtained for a simulation
using a point-like tip potential with $d_{\mathrm{tip}}=5$nm and
$V_{\mathrm{t}}=5E_{\mathrm{F}}$. }
\end{figure}

Another interesting and previously unexplored interference scenario
takes place in high mobility samples when a small number of hard impurities
is present. In Fig. \ref{fig:imp}(a-c) we show SGM images for a single
hard impurity within the device (with position marked by the black
dot) for temperatures $T=0$, 1K, and 4K. The characteristic quasi-elliptic
fringes visible in the SGM images can be explained as result of the
interference between electron waves following two different paths
between the QPC and the impurity: (i) a direct path of length $r_{\mathrm{q-i}}$
and (ii) a path of length $r_{\mathrm{q-t-i}}=r_{\mathrm{q-t}}+r_{\mathrm{t-i}}$
induced by the reflection on the depleted area below the tip (see
Fig. \ref{fig:elips}(a)). When the length difference is an integer
multiple of the Fermi wavelength, the interference is constructive at
the impurity location, resulting in a stronger backscattering and
a lower conductance. The resulting interference fringes can be approximated
as 
\begin{equation}
G\propto-\cos\left(k_{\mathrm{F}}\left(r_{\mathrm{q-t-i}}-r_{\mathrm{q-i}}\right)\right).\label{eq:ellips}
\end{equation}
The map calculated from Eq. \eqref{eq:ellips} is presented in Fig.
\ref{fig:elips}(b) and it can be compared with Fig. \ref{fig:elips}(c),
where we show the SGM image calculated for a point-like tip ($d_{\mathrm{tip}}=5$nm
and $V_{\mathrm{t}}=5E_{\mathrm{F}}$ such that $d_{\mathrm{delp}}=10$nm).
The white dashed lines in Fig. \ref{fig:elips}(b) and (c) represent
the isolines for $r_{\mathrm{q-t-i}}-r_{\mathrm{q-i}}=\lambda_{\mathrm{F}}/2$,
i.e. the position of the tip leading to the first destructive interference
between the paths marked in Fig. 8(a).

\begin{figure}
\begin{centering}
\includegraphics[width=1\columnwidth]{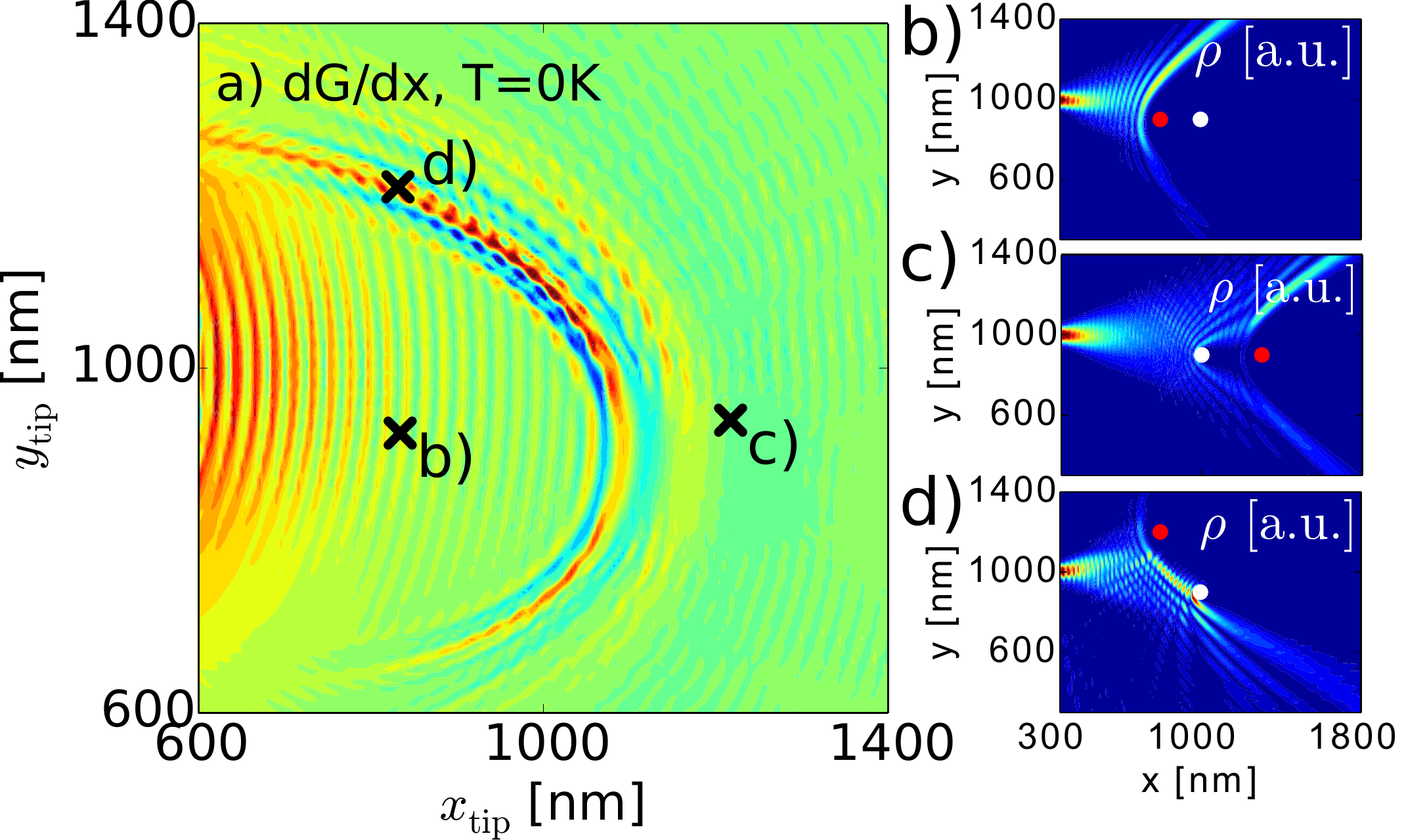} 
\par\end{centering}

\caption{\label{fig:esit}a) Same as in Fig. \ref{fig:imp}(a) but $dG/dx$
image. b-d) Scattering electron densities obtained for tip positions
denoted by crosses in (a). The white dot is the impurity and the red
dot is the tip position.}
\end{figure}

In Figs. \ref{fig:esit}(c-d) we show the scattering probability densities
that are at the origin of subsequent interference fringes visible
in SGM images in Fig. \ref{fig:esit}(a) or in Fig. \ref{fig:imp}(a).
The circular fringes visible inside the ellipse -- see Fig. \ref{fig:esit}(b)
-- are characteristic of a clean impurity-free sample and appear for
the impurity hidden by the tip depletion area {[}as in Fig. 5(d){]}.
On the other hand in Fig. \ref{fig:esit}(c) the tip is located in
the shadow of the impurity which results in strongly suppressed fringes,
since very small electron flow arrives to the tip and thus the conductance
map weakly depends on the tip position. In other tip positions {[}as
in Fig. 9(d){]} the process involves both the impurity and the tip
{[}cf. Fig. \ref{fig:elips}(a){]} producing the elliptic fringes.

From Fig. \ref{fig:imp}(c) one can see that the elliptic pattern
is thermally more stable than the circular fringes which decay rapidly
with the distance to the QPC. The most stable elliptic fringe is the
one for which the length difference between two paths in Fig. \ref{fig:elips}(a)
$r_{\mathrm{q-i}}-r_{\mathrm{q-t-i}}$ is equal to the half the Fermi
wavelength ($\lambda_{F}/2=27.35$ nm) -- which is much shorter than
the thermal length. 

In order to explain the exact position of the first elliptic fringe
one needs to account for the finite size of the depletion area and
the electron incidence angle $\alpha$ to the depletion area (see
Fig. 8(a)). Since the kinetic energy related to the electron motion
in the direction normal to the tip equipotential lines is $E_{F}\cos(\alpha)$,
the reflection point is located at a distance $d_{\mathrm{refl}}$
from the tip given by $E_{F}\cos(\alpha)=V_{\mathrm{tip}}(r=d_{\mathrm{refl}})=eV_{\mathrm{t}}d_{\mathrm{tip}}^{2}/\left(d_{\mathrm{tip}}^{2}+d_{\mathrm{depl}}^{2}\right)$.
From this condition one derives the reflection radius

\begin{align}
d_{\mathrm{refl}} & =d_{\mathrm{tip}}\sqrt{\frac{eV_{\mathrm{t}}}{E_{\mathrm{F}}\cos\left(\alpha\right)}-1}.\label{eq:depl}
\end{align}
which equals the depletion radius $d_{\mathrm{depl}}$ \eqref{eq:ddepl}
for the normal incidence ,$\alpha=0$ but is much larger for higher
incidence angles. In Fig. \ref{fig:imp}(a-c) three lines have been
drawn on the SGM images: (i) solid line is an ellipse corresponding
to the first interference fringe for a point-like tip potential and
denotes the first interference fringe (same as white dashed lines
in Fig. \ref{fig:elips}(b-c)); (ii) The central ellipse corresponds
to the smallest ellipse simply enlarged by $d_{\mathrm{depl}}$ in
the normal direction; and (iii) The largest contour corresponds to
the smallest ellipse but enlarged by $d_{\mathrm{refl}}$ from Eq.
\eqref{eq:depl} in the normal direction. This contour is no longer an ellipse
and we refer to this kind of curve by quasi-elliptic/ellipse (QE) in the following.
In order to fit this model to the SGM image we have set $d_{\mathrm{tip}}=75$nm in Eq. \eqref{eq:depl}
which is about the nominal value of 80nm. We have to move slightly
the impurity location by 20 nm to the left which is of the order of
the impurity radius. The idea of the incidence-angle-dependent penetration
depth was employed in a recent work of Ref. \cite{Steinacher2016}
in which the authors analyzed small-angle scattering trajectories
induced by potential barriers lower than the Fermi energy.

\begin{figure}
\begin{centering}
\includegraphics[width=1\columnwidth]{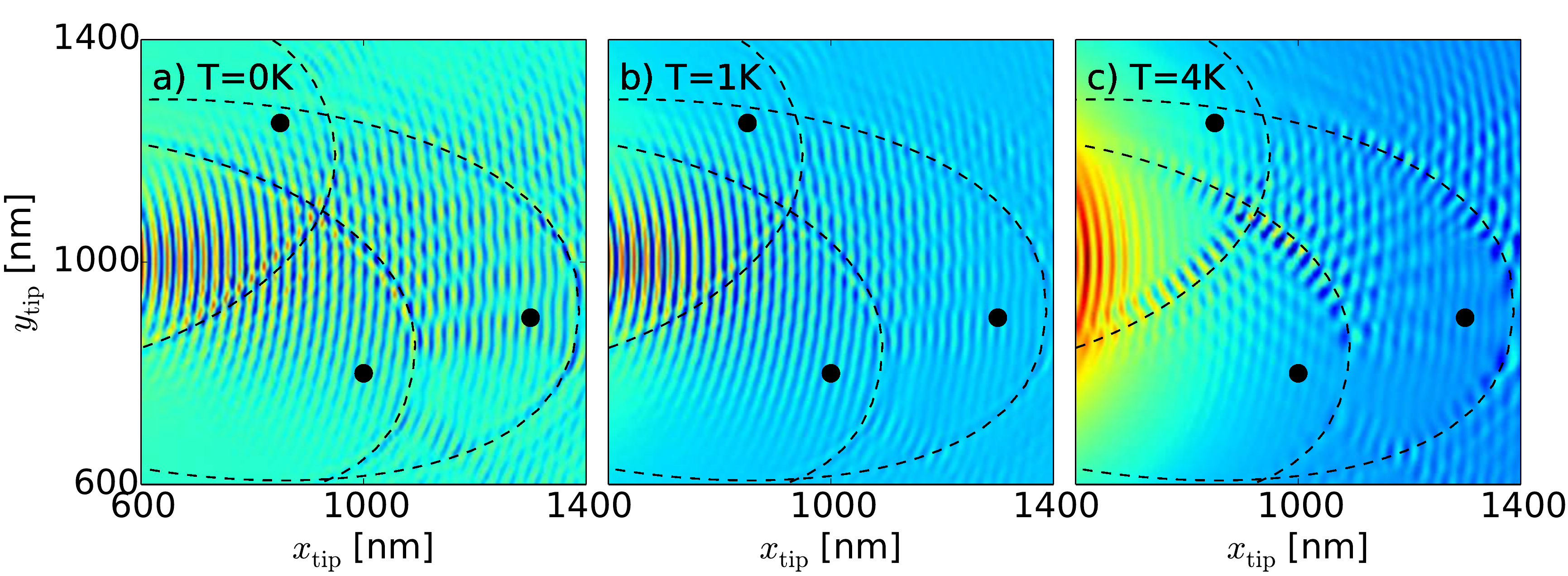} 
\par\end{centering}

\caption{\label{fig:ms}a-c) The $dG/dx$ SGM images (in arbitrary units) similar
like those in Fig. \ref{fig:imp}(a-c) but for three hard impurities
denoted by dots. Dashed lines present calculated position of the first
fringe from Eq. \eqref{eq:depl}.}
\end{figure}

Figure \ref{fig:ms}(a-c) show SGM images for three hard impurities
in the system with set of QE fringes. The dashed lines
show QEs obtained from Eq. \eqref{eq:depl} with $d_{\mathrm{tip}}=75$nm,
which agree with the value used in the simulation. Note that for a
few hard impurities, the SGM images resolve the QE
fringes resulting from separate interference scenarios.

\subsection{Experimental maps for hard scatterers}

\begin{figure}
\begin{centering}
\includegraphics[width=1\columnwidth]{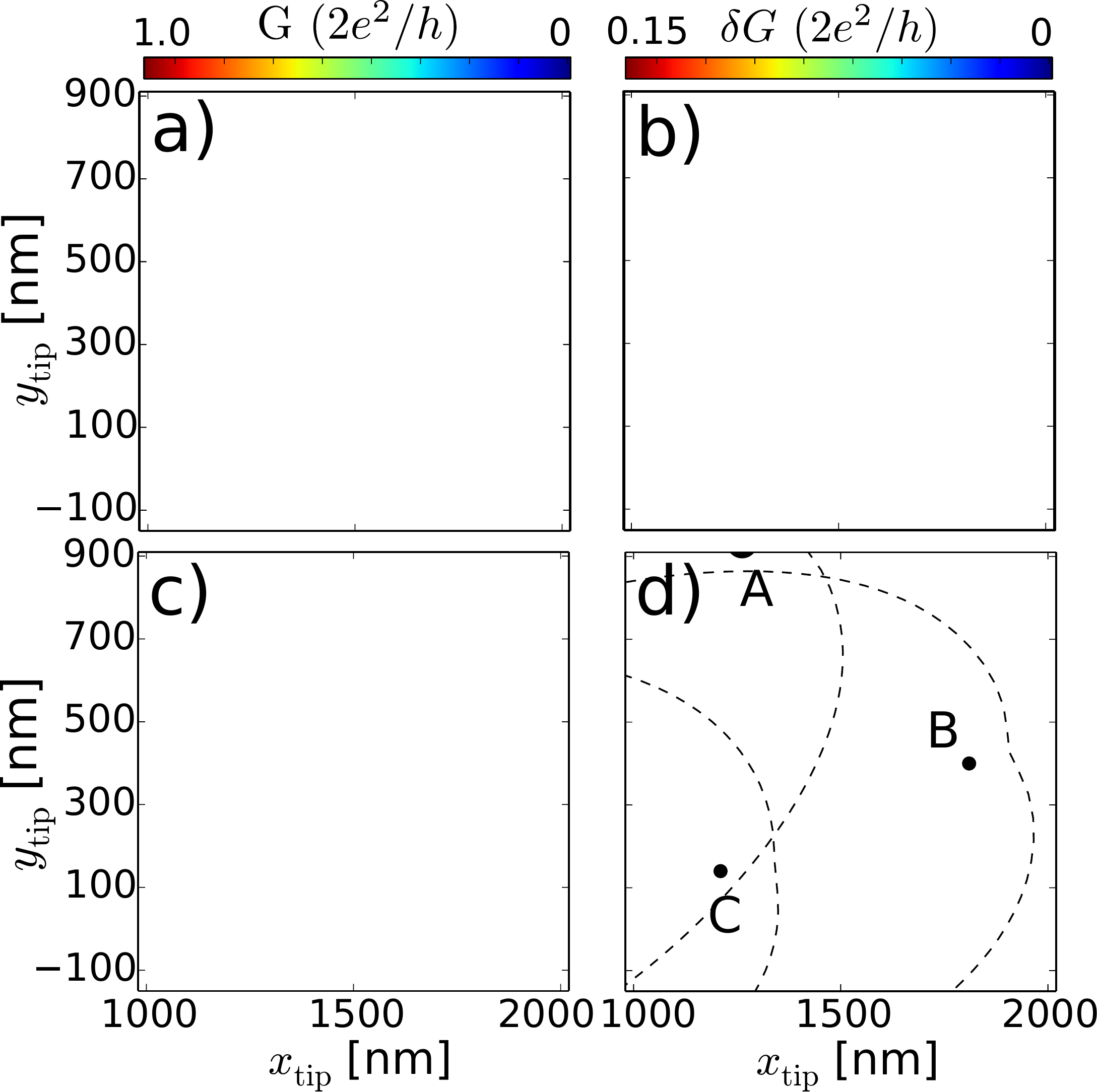} 
\par\end{centering}

\caption{\label{fig:exp}(a) Experimental SGM image of the conductance as a
function of the tip position for a tip voltage of $-6$ V. (b) SGM
image of the transconductance, i.e. the response of the conductance
to a 50 mV ac modulation applied to the tip around its dc value of
$-6$ V. The data plotted in (a) and (b) are measured simultaneously.
The origin of the coordinates is the center of QPC. (c) Numerical
derivative of (b) with respect to the horizontal axis, giving higher
contrast to the lines. (d) The dashed lines show the theoretical tip
position of the first fringe calculated for the three impurity positions
indicated by the dots (A,B,C) with Eq. \eqref{eq:depl} with $\frac{eV_{\mathrm{t}}}{E_{\mathrm{F}}}=1.2$
and $d_{\mathrm{tip}}=200$nm for points A,B and $\frac{eV_{\mathrm{t}}}{E_{\mathrm{F}}}=1.5$,
$d_{\mathrm{tip}}=170$nm for point C.}
\end{figure}

\begin{figure*}[t]
\centering{}\includegraphics[width=18cm]{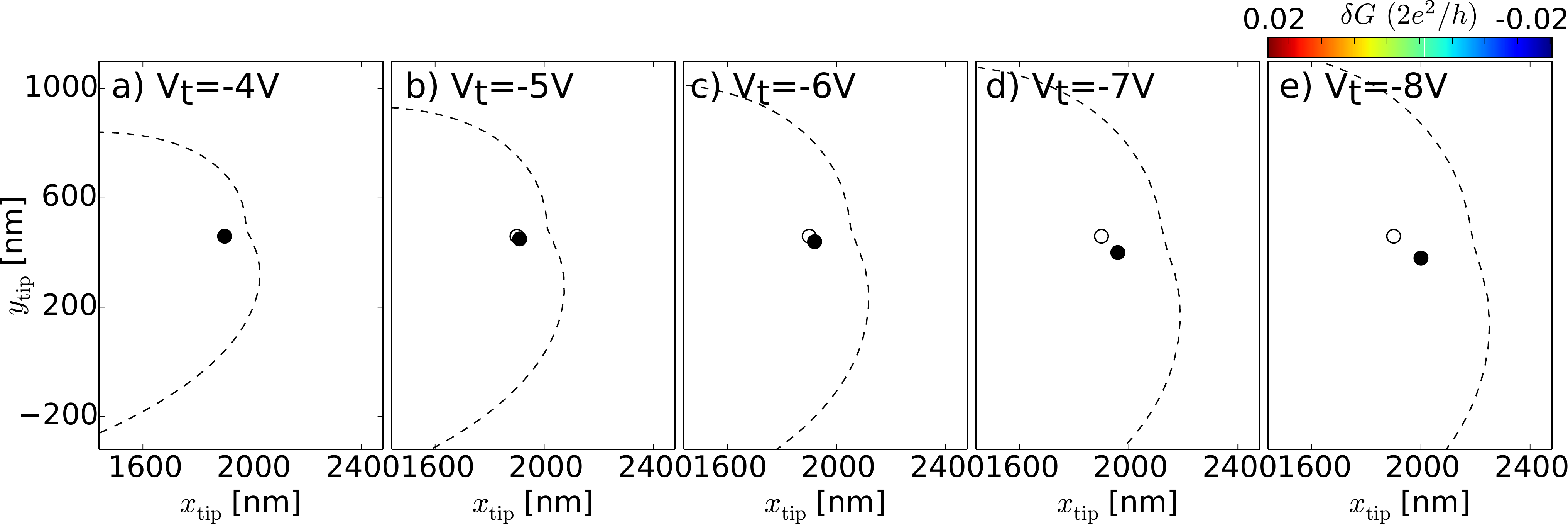} \caption{\label{figure2} Experimental SGM images of the transconductance for
different tip voltages : (a) $-4$ V, (b) $-5$ V, (c) $-6$ V, (d)
$-7$ V, (e) $-8$ V. In addition to the increasing gating effect,
the lines due to the presence of hard scatterers moves to the right
and change in curvature. Note that these images have been recorded
for a slightly smaller QPC transmission than in Fig.~\ref{fig:exp}
resulting in a faster closing of the QPC channel. The filled dots
show the assumed position of the impurity, while the empty dots indicate
the impurity position in (a). Dashed lines represent the theoretical
tip position of the first fringe calculated from Eq. \eqref{eq:depl}.}
\end{figure*}

For the experiment we use the same series of samples as in Refs. \cite{Brun2014,Brun2016}
for which the interference fringes between the QPC and the tip --
independently of the hard scatterers -- were reported previously at
low temperature. The presence of hard scatterers can be more easily
identified in SGM images recorded at a higher temperature, for which
the \textquotedbl{}clean\textquotedbl{} interference fringes disappear.
In this section, we illustrate the effects of single hard scatterers
in a high mobility sample by discussing a SGM experiment performed
at 4.2 K. The QPC is defined in a 2DEG located 105 nm below the surface
of a GaAs/AlGaAs heterostructure. The 2DEG has a $2.5\times10^{11}$
cm$^{-2}$ electron density and a $1.0\times10^{6}$ cm$^{2}$V$^{-1}$s$^{-1}$
electron mobility \cite{Brun2014,Brun2016}. The QPC is defined by
a Ti/Au split gate whose rectangular gap is 350 nm wide and 200 nm
long. The device is mounted in a cryogenic scanning probe microscope
and cooled down to a temperature of 4.2 K. The tip of the SGM microscope
is a commercial platinum-coated AFM tip fixed with silver epoxy to
a tuning fork which is used as the force sensor for topographic imaging.
In SGM mode, the tip is scanned above the 2DEG with a constant tip
voltage of $-6$ V and a tip-to-surface distance of 35 nm. The QPC
gate voltage is kept fixed at $-0.8$ V in order to have the QPC conductance
equal to one conductance quantum $2e^{2}/h$ when the tip is far from
the QPC. To enhance the sensitivity of the measurement to small tip-induced
effects, a small ac voltage modulation is applied to the tip and the
demodulated current response gives a transconductance signal. The
conductance $G=I^{{\rm ac}}/V_{{\rm bias}}^{{\rm ac}}$ is measured
with a 100 $\mu$V ac bias voltage applied between source and drain,
while the transconductance $\delta G=I^{{\rm ac}}/V_{{\rm bias}}^{{\rm dc}}$
is measured with a 50 mV ac voltage applied to the tip and a 150 $\mu$V
dc bias voltage applied between source and drain. The current flowing
through the QPC is amplified and the response to the ac excitation
is measured with a lock-in technique.

The SGM images plotted in Fig.~\ref{fig:exp}(a,b) present the conductance
and transconductance signals as a function of the tip position. The
center of the QPC is located at coordinates $(0,0)$ as determined
by SGM images recorded above the QPC and higher tip-to-surface distance
(data not shown). While the conductance image simply shows the gating
effect of the tip on the QPC transmission, the transconductance image
shows several additional lines. The origin of these lines is attributed
to the presence of hard scatterers in the 2DEG as discussed above
in Sec.~\ref{sub:hs}. 

In the following discussion, we assume that the lines arise from a
single hard scatterer, although we are aware of the possibility that
more impurities are involved. The dashed lines in Fig.~\ref{fig:exp}(d)
show QEs fitted using Eq.~\eqref{eq:depl} in which we
employ $\frac{V_{t}}{E_{F}}=1.2$ and $d_{\mathrm{tip}}=200$nm for
the impurities A and B. In order to fit Eq. \eqref{eq:depl} to fringes
originating from impurities A and B we set position of the QPC to
$r_{\mathrm{QPC}}=\left(50,-100\right)$nm, $r_{\mathrm{A}}=\left(1260,920\right)$nm
and $r_{\mathrm{B}}=\left(1810,400\right)$nm. Note, that the QPC
position obtained from the fit is shifted with respect to the center
of the QPC {[} nominally (0,0){]} which results from the fact that
the interference result from scattering between the tip and the QPC
gates -- and not the QPC entrance \cite{Jura2009} -- thus the QPC
focal point of the QE is not located at the entrance of
the QPC. At the scale of Fig. 11, the shift of $r_{\mathrm{QPC}}$
from the origin is small anyway. For the impurity C we get slightly
different values $\frac{V_{t}}{E_{F}}=1.5$, $d_{\mathrm{tip}}=170$nm,
$r_{\mathrm{QPC}}=\left(0,50\right)$ nm and $r_{\mathrm{B}}=\left(1210,140\right)$
nm. The difference in the tip potential parameters may be due to the
screening of the tip by the gates ($C$ is closer to the gates than
$A$ and $B$). The observed number of impurities in the scanned area
$1\mu m\times1\mu m$ gives impurity density $n_{\mathrm{imp}}\approx3${[}$\frac{1}{\mu m^{2}}${]},
which can be used to roughly estimate the electron mobility inside
the 2DEG with semiclassical formula $\mu=\frac{ql_{p}}{\sqrt{2mE_{\mathrm{F}}}}=\frac{q\lambda_{\mathrm{F}}l_{p}}{h}$,
where $l_{p}=1/n_{\mathrm{imp}}\lambda_{\mathrm{F}}$ is the mean
free path. The value of $l_{p}$ is estimated from the semi-classical
Broglie's assumption of electron being a particle of diameter $\lambda_{\mathrm{F}}$
colliding with point like scatters uniformly distributed in the sample.
The approximated expression for electron mobility reduces then to
simple formula $\mu=\frac{q}{n_{\mathrm{imp}}h}\approx$$0.8\times10^{6}$
cm$^{2}$V$^{-1}$s$^{-1}$ .


The evolution of the lines with the tip voltage is presented in Fig.~\ref{figure2}(a-e).
When the tip voltage is made more negative, the lines move to larger
distances from the QPC and become wider in the transverse direction
(smaller curvature). This behavior is consistent with the simulations
presented in Fig.~\ref{fig:imp}(c), where a larger depletion disk
below the tip results in a lower eccentricity of the QE
lines. The dashed lines in Figs. \ref{figure2} show the results of
Eq. \eqref{eq:depl} that are obtained with increasing values of $d_{\mathrm{tip}}=\left\{ 165,205,235,260,280\right\} $nm
and ratio $\frac{V_{\mathrm{t}}}{E_{\mathrm{F}}}=\left\{ 1.19,1.22,1.3,1.4,1.45\right\} $,
respectively. This non-trivial evolution of the tip-induced potential
parameters (non constant $d_{\mathrm{tip}}$ and slowly varying $V_{\mathrm{t}}$)
reflects the complex behavior of the non-linear screening in case
of partial depletion. We obtain a change of the tip radius to be about
$\sim30$ nm for a 1 V change on the tip. In order to obtain a good
fit between the first QE lines and the analytical expression,
we shift the positions of the impurity in Fig. \ref{figure2}(b-e)
(filled circles) with respect to the calculated position in the first
image (a) (empty circles) by about (20nm,-20nm) per image. The reason
of this shift may be caused by the drift of the sample with respect
to the tip position due to the long acquisition time of 2$\frac{1}{2}$h
per image.

\section{Summary and Conclusions}

To summarize we have discussed the role of smooth and hard impurities
in the 2DEG on the SGM images. We have shown that the funnel like
features that appear in conductance maps result from the splitting
of the branches by hard impurities, and the position of the impurity
is always shifted in the SGM images due the finite size of the depletion
area. We have shown that 1D interpretation of branches can be used
to explain most of the features present in the conductance maps, including
their thermal stability. Additionally, we have discussed that in presence
of a small number of hard impurities in high mobility sample characteristic
quasi elliptic fringes can be found in the SGM images even at reasonably
high temperatures $\sim4K$. We have explained those findings in terms
of interference between two paths involving both the tip and the impurity
with  length difference of the order of $\lambda_{\mathrm{F}}/2$.
We have provided an experimental evidence for this interference processes
as well as a simple analytical formula which can be used to extract
the position of the impurity and to estimate the depletion radius
due to the tip.

\subsection*{Acknowledgments }

This work was supported by National Science Centre according to decision
DEC- 2015/17/N/ST3/02266, and by PL-Grid Infrastructure. The first
author is supported by the scholarship of Krakow Smoluchowski Scientific
Consortium from the funding for National Leading Reserch Centre by
Ministry of Science and Higher Education (Poland) and by the Etiuda
stipend of the National Science Centre (NCN) according to decision
DEC-2015/16/T/ST3/00310. The experimental work was supported by the
French Agence Nationale de la Recherche (``ITEM-exp'' project) and
the sample fabrication was done by U. Gennser and D. Mailly from CNRS/LPN.

\section*{Appendix}

\subsection{Description of the numerical method}

We start from the derivation of the scattering boundary conditions
using the approach from Ref. \cite{Kirkner1990}. Let us assume that
the simulated device can be approximated by tight-binding like Hamiltonian
$\boldsymbol{H}$, in our case such Hamiltonian is generated from
finite difference approximation of the derivatives of the differential
operators in $\boldsymbol{H}$ \cite{Kolasinski2016Lande}. Additionally,
we follow the Ref. \cite{Zwierzycki2008} and we divide the system
into consecutive slices $\boldsymbol{H}_{\mathrm{i}}$ connected by
coupling matrices $\boldsymbol{\tau}_{\mathrm{i}}$, forming block-tridiagonal
systems of linear equations for the scattering wave function $\boldsymbol{c}$
inside the system 
\[
-\boldsymbol{\tau}_{\mathrm{i}-1}\boldsymbol{c}_{\mathrm{i-1}}+\left(E_{\mathrm{F}}-\boldsymbol{H}_{\mathrm{i}}\right)\boldsymbol{c}_{\mathrm{i}}-\boldsymbol{\tau}_{\mathrm{i}}^{\dagger}\boldsymbol{c}_{\mathrm{i+1}}=0.
\]
In the lead region (semi infinite lead can be located in any part
of the system) the system is assumed to be homogeneous thus one may
drop the indices in the matrices and write 
\begin{equation}
-\boldsymbol{\tau}\boldsymbol{c}_{\mathrm{i-1}}+\left(E_{\mathrm{F}}-\boldsymbol{H}\right)\boldsymbol{c}_{\mathrm{i}}-\boldsymbol{\tau}^{\dagger}\boldsymbol{c}_{\mathrm{i+1}}=0,\label{eq:Hc}
\end{equation}
 which can be solved by Bloch substitution $\boldsymbol{c}_{\mathrm{n}}=\lambda^{n}\boldsymbol{u}$
\cite{Zwierzycki2008} leading to quadratic eigenvalue equation for
the transverse modes \cite{Zwierzycki2008,AndoWFM1991}
\[
-\boldsymbol{\tau}\boldsymbol{u}+\lambda\left(E_{\mathrm{F}}-\boldsymbol{H}\right)\boldsymbol{u}-\lambda^{2}\boldsymbol{\tau}^{\dagger}\boldsymbol{u}=0,
\]
 which can be transformed to generalized eigenvalue problem (GEP)
of double size 
\begin{equation}
\left(\begin{array}{cc}
\boldsymbol{0} & \boldsymbol{1}\\
-\boldsymbol{\tau} & E_{\mathrm{F}}-\boldsymbol{H}
\end{array}\right)\left(\begin{array}{c}
\boldsymbol{u}\\
\lambda\boldsymbol{u}
\end{array}\right)=\lambda\left(\begin{array}{cc}
\boldsymbol{1} & \boldsymbol{0}\\
\boldsymbol{0} & \boldsymbol{\tau}^{\dagger}
\end{array}\right)\left(\begin{array}{c}
\boldsymbol{u}\\
\lambda\boldsymbol{u}
\end{array}\right).\label{eq:modes}
\end{equation}
 We solve it numerically by converting it to a standard eigenvalue
problem (SEP), since in our case $\boldsymbol{\tau}^{\dagger}$ is
invertible. If $\boldsymbol{\tau}^{\dagger}$ is non-invertible or
ill-conditioned one may use more sophisticated methods which incorporate
Singular Value Decomposition (SVD) of $\boldsymbol{\tau}^{\dagger}$
matrix \cite{Rungger2008}.

The eigenvalues of equation above are then grouped into incoming $\left\{ \lambda_{m,+},\boldsymbol{u}_{m,+}\right\} $
and outgoing modes $\left\{ \lambda_{m,-},\boldsymbol{u}_{m,-}\right\} $,
with each propagating mode $\boldsymbol{u}_{m,\pm}$ (i.e. with $\left|\lambda_{m,\pm}\right|=1$)
normalized to carry the unit value of quantum flux. For more detailed
description see \cite{Zwierzycki2008}.

The solution in the semi-infinite lead for the $m$-th incoming mode
can be expressed in terms of superposition transverse modes 
\begin{equation}
\boldsymbol{c}_{\mathrm{n}}=\lambda_{m,+}^{n}\boldsymbol{u}_{m,+}+\sum_{k=1}^{N}r_{\mathrm{k}}\lambda_{k,-}^{n}\boldsymbol{u}_{k,-},\label{eq:c0full}
\end{equation}
where $\lambda_{m,+}^{n}$is the Bloch factor \cite{Zwierzycki2008}
for the $m$-th incoming mode $\boldsymbol{u}_{m,+}$ and $N$ is
the number of sites in the lead slice i.e. the size of the vector
$\boldsymbol{u}_{\mathrm{k,\pm}}$. Vector $\boldsymbol{u}_{k,-}$
denotes the $k$-th outgoing mode. For more detailed description how
the transverse modes are calculated see Ref. \cite{Zwierzycki2008}.
By choosing the frame of coordinates such that $i=0$ denotes the
first slice in the considered system one may expand the wave function
at this slice in terms of transverse modes 
\begin{equation}
\boldsymbol{c}_{\mathrm{0}}=\boldsymbol{u}_{m,+}+\sum_{k=1}^{N}r_{\mathrm{k}}\boldsymbol{u}_{k,-},\label{eq:c0}
\end{equation}
 note that $\boldsymbol{c}_{\mathrm{0}}$ is now a part of discretized
system. By projecting $\bra{\boldsymbol{u}_{p,-}}$ on Eq. \eqref{eq:c0},
we get 
\[
\braket{\boldsymbol{u}_{p,-}|\boldsymbol{c}_{\mathrm{0}}}=\braket{\boldsymbol{u}_{p,-}|\boldsymbol{u}_{m,+}}+\sum_{k=1}^{N}r_{\mathrm{k}}\braket{\boldsymbol{u}_{p,-}|\boldsymbol{u}_{k,-}},
\]
 with $\braket{a|b}=\sum_{k}^{N}a^{*}(k)b(k)$, which can be written
in terms of matrices 
\[
\boldsymbol{r}=\boldsymbol{S}\left(\boldsymbol{Q}-\boldsymbol{B}_{m}\right),
\]
with $\boldsymbol{r}=\left\{ r_{\mathrm{1}},r_{\mathrm{2}}\ldots,r_{N}\right\} $,
$\boldsymbol{Q}_{\mathrm{p}}=\braket{\boldsymbol{u}_{p,-}|\boldsymbol{c}_{\mathrm{0}}}$,
$\boldsymbol{B}_{m,p}=\braket{\boldsymbol{u}_{p,-}|\boldsymbol{u}_{m,+}}$
and $S_{\mathrm{p,k}}^{\mathrm{-1}}=\braket{\boldsymbol{u}_{p,-}|\boldsymbol{u}_{k,-}}$.
Additionally, by forcing the derivative of the wave function to be
continuous at the device boundary we obtain second condition 
\[
\boldsymbol{c}_{\mathrm{0}}-\boldsymbol{c}_{\mathrm{-1}}=\left(1-\lambda_{m,+}^{-1}\right)\boldsymbol{u}_{m,+}+\sum_{k=1}^{N}\left(1-\lambda_{k,-}^{-1}\right)r_{\mathrm{k}}\boldsymbol{u}_{k,-}.
\]
 Then by substituting the $\boldsymbol{r}$ vector to the equation
above we get 
\begin{align*}
\boldsymbol{c}_{\mathrm{0}}-\boldsymbol{c}_{\mathrm{-1}} & =\left(1-\lambda_{m,+}^{-1}\right)\boldsymbol{u}_{m,+}.\\
 & +\sum_{k,p=1}^{N}\left(1-\lambda_{k,-}^{-1}\right)S_{\mathrm{k,p}}\left(Q_{\mathrm{p}}-B_{m,p}\right)\boldsymbol{u}_{k,-}.
\end{align*}
 Let us now simplify the expression above, by starting from the first
term in the sum on the right side 
\begin{align*}
\sum_{k,p=1}^{N}\left(1-\lambda_{k,-}^{-1}\right)S_{\mathrm{k,p}}Q_{\mathrm{p}}\boldsymbol{u}_{k,-} & =\\
=\sum_{k,p=1}^{N}\left(1-\lambda_{k,-}^{-1}\right)S_{\mathrm{k,p}}\braket{\boldsymbol{u}_{p,-}|\boldsymbol{c}_{\mathrm{0}}}\boldsymbol{u}_{k,-} & =\\
=\sum_{k,p=1}^{N}\left(1-\lambda_{k,-}^{-1}\right)S_{\mathrm{k,p}}\sum_{i}^{N}u_{p,-}^{*}\left(i\right)c_{\mathrm{0}}\left(i\right)\boldsymbol{u}_{k,-} & =\\
=\sum_{i,k,p=1}^{N}\boldsymbol{u}_{k,-}\left(1-\lambda_{k,-}^{-1}\right)S_{\mathrm{k,p}}u_{p,-}^{*}\left(i\right)c_{\mathrm{0}}\left(i\right) & =\\
=\boldsymbol{U}_{-}\left(\boldsymbol{1}-\boldsymbol{\Lambda}_{-}^{-1}\right)\boldsymbol{S}\boldsymbol{U}_{-}^{\dagger}\boldsymbol{c}_{\mathrm{0}}.
\end{align*}
Where columns of matrix $\boldsymbol{U}_{\pm}$ are constructed from
transverse modes $\boldsymbol{U}_{\pm}=\left(\ket{\boldsymbol{u}_{1,\pm}},\ket{\boldsymbol{u}_{2,\pm}},\ldots,\ket{\boldsymbol{u}_{\mathrm{N},\pm}}\right)$
and $\Lambda_{i,j,\pm}=\delta_{i,j}\lambda_{i,\pm}$. Analogically
for the second term, we obtain 
\begin{align*}
\sum_{k,p=1}^{N}\left(1-\lambda_{k,-}^{-1}\right)S_{\mathrm{k,p}}B_{m,p}\boldsymbol{u}_{k,-} & =\boldsymbol{U}_{-}\left(\boldsymbol{1}-\boldsymbol{\Lambda}_{-}^{-1}\right)\boldsymbol{S}\boldsymbol{U}_{-}^{\dagger}\boldsymbol{u}_{m,+}.
\end{align*}
 Thus we get 
\begin{align}
\boldsymbol{c}_{\mathrm{0}}-\boldsymbol{c}_{\mathrm{-1}} & =\left(1-\lambda_{m,+}^{-1}\right)\boldsymbol{u}_{m,+}+\nonumber \\
+ & \boldsymbol{U}_{-}\left(\boldsymbol{1}-\boldsymbol{\Lambda}_{-}^{-1}\right)\boldsymbol{S}\boldsymbol{U}_{-}^{\dagger}\boldsymbol{c}_{\mathrm{0}}-\boldsymbol{U}_{-}\left(\boldsymbol{1}-\boldsymbol{\Lambda}_{-}^{-1}\right)\boldsymbol{S}\boldsymbol{U}_{-}^{\dagger}\boldsymbol{u}_{m,+}.\label{eq:c02}
\end{align}
 The expression above can be further simplified by noticing that $\boldsymbol{S}=\left(\boldsymbol{U}_{-}^{\dagger}\boldsymbol{U}_{-}\right)^{-1}=\boldsymbol{U}_{-}^{-1}\left(\boldsymbol{U}_{-}^{\dagger}\right)^{-1}$,
hence the matrix 
\begin{align*}
\boldsymbol{U}_{-}\left(\boldsymbol{1}-\boldsymbol{\Lambda}_{-}^{-1}\right)\boldsymbol{S}\boldsymbol{U}_{-}^{\dagger} & =\boldsymbol{U}_{-}\left(\boldsymbol{1}-\boldsymbol{\Lambda}_{-}^{-1}\right)\boldsymbol{U}_{-}^{-1}\left(\boldsymbol{U}_{-}^{\dagger}\right)^{-1}\boldsymbol{U}_{-}^{\dagger}\\
 & =\boldsymbol{1}-\boldsymbol{U}_{-}\boldsymbol{\Lambda}_{-}^{-1}\boldsymbol{U}_{-}^{-1}\equiv\boldsymbol{1}-\boldsymbol{F_{-}}.
\end{align*}
 The matrix $\boldsymbol{F}_{\pm}\equiv\boldsymbol{U}_{\pm}\left(\boldsymbol{U}_{\pm}\boldsymbol{\Lambda}_{\pm}\right)^{-1}$
is the Bloch matrix. The final formula for Eq. \eqref{eq:c02} is
then 
\[
\boldsymbol{c}_{\mathrm{-1}}=\boldsymbol{F}_{-}\boldsymbol{c}_{\mathrm{0}}+\left(\lambda_{m,+}^{-1}\boldsymbol{1}-\boldsymbol{F}_{-}\right)\boldsymbol{u}_{m,+}.
\]
By inserting this to the Hamiltonian \eqref{eq:Hc} for $i=0$ one
removes the dependence of the $\boldsymbol{c}_{\mathrm{-1}}$ slice
from the linear system which gives 
\[
\left(E_{\mathrm{F}}-\boldsymbol{H}-\boldsymbol{\tau}\boldsymbol{F}_{-}\right)\boldsymbol{c}_{\mathrm{0}}-\boldsymbol{\tau}^{\dagger}\boldsymbol{c}_{\mathrm{1}}=\boldsymbol{\tau}\left(\lambda_{m,+}^{-1}\boldsymbol{1}-\boldsymbol{F}_{-}\right)\boldsymbol{u}_{m,+}.
\]
 We note that 
\begin{align*}
\boldsymbol{F}_{+}\ket{\boldsymbol{u}_{m,+}} & =\boldsymbol{U}_{+}\boldsymbol{\Lambda}_{+}^{-1}\boldsymbol{U}_{+}^{-1}\ket{\boldsymbol{u}_{m,+}}=\\
 & =\boldsymbol{U}_{+}\ket{0,\ldots,\lambda_{m,+}^{-1},\ldots,0}=\\
 & =\lambda_{m,+}^{-1}\ket{\boldsymbol{u}_{m,+}},
\end{align*}
 hence the right side can be written in a more compact form 
\begin{align*}
\boldsymbol{\tau}\left(\lambda_{m,+}^{-1}\boldsymbol{1}-\boldsymbol{F}_{-}\right)\ket{\boldsymbol{u}_{m,+}} & =\boldsymbol{\tau}\left(\boldsymbol{F}_{+}-\boldsymbol{F}_{-}\right)\ket{\boldsymbol{u}_{m,+}}\equiv\Gamma_{\mathrm{m}},
\end{align*}
 To summarize the system of linear equation for the case of two terminal
device can be written in the following way 
\begin{align}
\left(E_{\mathrm{F}}-\left(\boldsymbol{H}_{0}+\boldsymbol{\Sigma}_{\mathrm{0}}\right)\right)\boldsymbol{c}_{\mathrm{0}}-\boldsymbol{\tau}_{0}^{\dagger}\boldsymbol{c}_{\mathrm{2}} & =\boldsymbol{\Gamma}_{0,\mathrm{m}},\label{eq:WFMa}\\
-\boldsymbol{\tau}_{\mathrm{i-1}}\boldsymbol{c}_{\mathrm{i-1}}+\left(E_{\mathrm{F}}-\boldsymbol{H}_{\mathrm{i}}\right)\boldsymbol{c}_{\mathrm{i}}-\boldsymbol{\tau}_{\mathrm{i}}^{\dagger}\boldsymbol{c}_{\mathrm{i+1}} & =0,\,\mathrm{for}\,0<i<\mathrm{N}\label{eq:WFMb}\\
\left(E_{\mathrm{F}}-\left(\boldsymbol{H}_{\mathrm{N}}+\boldsymbol{\Sigma}_{\mathrm{N}}\right)\right)\boldsymbol{c}_{\mathrm{N}}-\boldsymbol{\tau}_{\mathrm{N-1}}\boldsymbol{c}_{\mathrm{N-1}} & =0,\label{eq:WFMc}
\end{align}
where $\Sigma_{\mathrm{0/N}}=\boldsymbol{\tau}\boldsymbol{F}_{-}$
is the self energy calculated for left/right lead. Note, that in order
to obtain open boundary conditions we use the approach from the Quantum
Transmitting Boundary Method (QTBM) introduced in Ref. \cite{Kirkner1990},
but at the end we finish with the Wave Function Matching (WFM) equations
\cite{Zwierzycki2008}, which shows that both methods are algebraically
equivalent.

After solution of the scattering problem for a given $m$-th incoming
mode one may calculate transmission amplitudes from 
\begin{equation}
\boldsymbol{t}_{\mathrm{m}}=\boldsymbol{U}_{\mathrm{N},-}^{-1}\boldsymbol{c}_{\mathrm{N,m}},\label{eq:tm}
\end{equation}
 and reflection amplitudes as 
\begin{equation}
\boldsymbol{r}_{\mathrm{m}}=\boldsymbol{U}_{\mathrm{0},-}^{-1}\left(\boldsymbol{c}_{\mathrm{0,m}}-\boldsymbol{u}_{m,+}\right),\label{eq:rm}
\end{equation}
 with $\boldsymbol{U}_{\mathrm{0},-}$ and $\boldsymbol{U}_{\mathrm{N},-}$
being the outgoing modes matrices for input and output lead, respectively.

\bibliographystyle{apsrev4-1-nourl}
\bibliography{referencje}

\begin{thebibliography}{10}%
\makeatletter
\providecommand \@ifxundefined [1]{%
 \ifx #1\undefined \expandafter \@firstoftwo
 \else \expandafter \@secondoftwo
\fi
}%
\providecommand \@ifnum [1]{%
 \ifnum #1\expandafter \@firstoftwo
 \else \expandafter \@secondoftwo
\fi
}%
\providecommand \enquote [1]{``#1''}%
\providecommand \bibnamefont  [1]{#1}%
\providecommand \bibfnamefont [1]{#1}%
\providecommand \citenamefont [1]{#1}%
\providecommand\href[0]{\@sanitize\@href}%
\providecommand\@href[1]{\endgroup\@@startlink{#1}\endgroup\@@href}%
\providecommand\@@href[1]{#1\@@endlink}%
\providecommand \@sanitize [0]{\begingroup\catcode`\&12\catcode`\#12\relax}%
\@ifxundefined \pdfoutput {\@firstoftwo}{%
 \@ifnum{\z@=\pdfoutput}{\@firstoftwo}{\@secondoftwo}%
}{%
 \providecommand\@@startlink[1]{\leavevmode\special{html:<a href="#1">}}%
 \providecommand\@@endlink[0]{\special{html:</a>}}%
}{%
 \providecommand\@@startlink[1]{%
  \leavevmode
  \pdfstartlink
   attr{/Border[0 0 1 ]/H/I/C[0 1 1]}%
   user{/Subtype/Link/A<</Type/Action/S/URI/URI(#1)>>}%
  \relax
 }%
 \providecommand\@@endlink[0]{\pdfendlink}%
}%
\providecommand \url  [0]{\begingroup\@sanitize \@url }%
\providecommand \@url [1]{\endgroup\@href {#1}{\urlprefix}}%
\providecommand \urlprefix [0]{URL }%
\providecommand \Eprint[0]{\href }%
\@ifxundefined \urlstyle {%
  \providecommand \doi [1]{doi:\discretionary{}{}{}#1}%
}{%
  \providecommand \doi [0]{doi:\discretionary{}{}{}\begingroup
  \urlstyle{rm}\Url }%
}%
\providecommand \doibase [0]{http://dx.doi.org/}%
\providecommand \Doi[1]{\href{\doibase#1}}%
\providecommand \bibAnnote [3]{%
  \BibitemShut{#1}%
  \begin{quotation}\noindent
    \textsc{Key:}\ #2\\\textsc{Annotation:}\ #3%
  \end{quotation}%
}%
\providecommand \bibAnnoteFile [2]{%
  \IfFileExists{#2}{\bibAnnote {#1} {#2} {\input{#2}}}{}%
}%
\providecommand \typeout [0]{\immediate \write \m@ne }%
\providecommand \selectlanguage [0]{\@gobble}%
\providecommand \bibinfo [0]{\@secondoftwo}%
\providecommand \bibfield [0]{\@secondoftwo}%
\providecommand \translation [1]{[#1]}%
\providecommand \BibitemOpen[0]{}%
\providecommand \bibitemStop [0]{}%
\providecommand \bibitemNoStop [0]{.\EOS\space}%
\providecommand \EOS [0]{\spacefactor3000\relax}%
\providecommand \BibitemShut [1]{\csname bibitem#1\endcsname}%
\bibitem{Eriksson1996}%
  \BibitemOpen
  \bibfield{author}{%
  \bibinfo {author} {\bibfnamefont{M.~A.}\ \bibnamefont{Eriksson}}, \bibinfo
  {author} {\bibfnamefont{R.~G.}\ \bibnamefont{Beck}}, \bibinfo {author}
  {\bibfnamefont{M.}~\bibnamefont{Topinka}}, \bibinfo {author}
  {\bibfnamefont{J.~A.}\ \bibnamefont{Katine}}, \bibinfo {author}
  {\bibfnamefont{R.~M.}\ \bibnamefont{Westervelt}}, \bibinfo {author}
  {\bibfnamefont{K.~L.}\ \bibnamefont{Campman}},\ and\ \bibinfo {author}
  {\bibfnamefont{A.~C.}\ \bibnamefont{Gossard}},\ }%
  \bibfield{journal}{%
  \Doi{http://dx.doi.org/10.1063/1.117801}{\bibinfo {journal} {Applied Physics
  Letters}}\ }%
  \textbf{\bibinfo {volume} {69}},\ \bibinfo {pages} {671} (\bibinfo {year}
  {1996})%
  \bibAnnoteFile{NoStop}{Eriksson1996}%
\bibitem{Sellier2011}%
  \BibitemOpen
  \bibfield{author}{%
  \bibinfo {author} {\bibfnamefont{H.}~\bibnamefont{Sellier}}, \bibinfo
  {author} {\bibfnamefont{B.}~\bibnamefont{Hackens}}, \bibinfo {author}
  {\bibfnamefont{M.~G.}\ \bibnamefont{Pala}}, \bibinfo {author}
  {\bibfnamefont{F.}~\bibnamefont{Martins}}, \bibinfo {author}
  {\bibfnamefont{S.}~\bibnamefont{Baltazar}}, \bibinfo {author}
  {\bibfnamefont{X.}~\bibnamefont{Wallart}}, \bibinfo {author}
  {\bibfnamefont{L.}~\bibnamefont{Desplanque}}, \bibinfo {author}
  {\bibfnamefont{V.}~\bibnamefont{Bayot}},\ and\ \bibinfo {author}
  {\bibfnamefont{S.}~\bibnamefont{Huant}},\ }%
  \bibfield{journal}{%
  \bibinfo {journal} {Semicond. Sci. Technol.}\ }%
  \textbf{\bibinfo {volume} {26}},\ \bibinfo {pages} {064008} (\bibinfo {year}
  {2011})%
  \bibAnnoteFile{NoStop}{Sellier2011}%
\bibitem{Ferry2011}%
  \BibitemOpen
  \bibfield{author}{%
  \bibinfo {author} {\bibfnamefont{D.~K.}\ \bibnamefont{Ferry}}, \bibinfo
  {author} {\bibfnamefont{A.~M.}\ \bibnamefont{Burke}}, \bibinfo {author}
  {\bibfnamefont{R.}~\bibnamefont{Akis}}, \bibinfo {author}
  {\bibfnamefont{R.}~\bibnamefont{Brunner}}, \bibinfo {author}
  {\bibfnamefont{T.~E.}\ \bibnamefont{Day}}, \bibinfo {author}
  {\bibfnamefont{R.}~\bibnamefont{Meisels}}, \bibinfo {author}
  {\bibfnamefont{F.}~\bibnamefont{Kuchar}}, \bibinfo {author}
  {\bibfnamefont{J.~P.}\ \bibnamefont{Bird}},\ and\ \bibinfo {author}
  {\bibfnamefont{B.~R.}\ \bibnamefont{Bennett}},\ }%
  \bibfield{journal}{%
  \bibinfo {journal} {Sem. Sci. Tech.}\ }%
  \textbf{\bibinfo {volume} {26}},\ \bibinfo {pages} {043001} (\bibinfo {year}
  {2011})%
  \bibAnnoteFile{NoStop}{Ferry2011}%
\bibitem{Brun2014}%
  \BibitemOpen
  \bibfield{author}{%
  \bibinfo {author} {\bibfnamefont{B.}~\bibnamefont{Brun}}, \bibinfo {author}
  {\bibfnamefont{F.}~\bibnamefont{Martins}}, \bibinfo {author}
  {\bibfnamefont{S.}~\bibnamefont{Faniel}}, \bibinfo {author}
  {\bibfnamefont{B.}~\bibnamefont{Hackens}}, \bibinfo {author}
  {\bibfnamefont{G.}~\bibnamefont{Bachelier}}, \bibinfo {author}
  {\bibfnamefont{A.}~\bibnamefont{Cavanna}}, \bibinfo {author}
  {\bibfnamefont{C.}~\bibnamefont{Ulysse}}, \bibinfo {author}
  {\bibfnamefont{A.}~\bibnamefont{Ouerghi}}, \bibinfo {author}
  {\bibfnamefont{U.}~\bibnamefont{Gennser}}, \bibinfo {author}
  {\bibfnamefont{D.}~\bibnamefont{Mailly}}, \bibinfo {author}
  {\bibfnamefont{S.}~\bibnamefont{Huant}}, \bibinfo {author}
  {\bibfnamefont{V.}~\bibnamefont{Bayot}}, \bibinfo {author}
  {\bibfnamefont{M.}~\bibnamefont{Sanquer}},\ and\ \bibinfo {author}
  {\bibfnamefont{H.}~\bibnamefont{Sellier}},\ }%
  \bibfield{journal}{%
  \bibinfo {journal} {Nat Commun}\ }%
  \textbf{\bibinfo {volume} {5}} (\bibinfo {year} {2014}),\ \bibinfo {note}
  {article}%
  \bibAnnoteFile{NoStop}{Brun2014}%
\bibitem{Wees1988}%
  \BibitemOpen
  \bibfield{author}{%
  \bibinfo {author} {\bibfnamefont{B.~J.}\ \bibnamefont{van Wees}}, \bibinfo
  {author} {\bibfnamefont{H.}~\bibnamefont{van Houten}}, \bibinfo {author}
  {\bibfnamefont{C.~W.~J.}\ \bibnamefont{Beenakker}}, \bibinfo {author}
  {\bibfnamefont{J.~G.}\ \bibnamefont{Williamson}}, \bibinfo {author}
  {\bibfnamefont{L.~P.}\ \bibnamefont{Kouwenhoven}}, \bibinfo {author}
  {\bibfnamefont{D.}~\bibnamefont{van~der Marel}},\ and\ \bibinfo {author}
  {\bibfnamefont{C.~T.}\ \bibnamefont{Foxon}},\ }%
  \bibfield{journal}{%
  \Doi{10.1103/PhysRevLett.60.848}{\bibinfo {journal} {Phys. Rev. Lett.}}\ }%
  \textbf{\bibinfo {volume} {60}},\ \bibinfo {pages} {848} (\bibinfo {month}
  {Feb}\ \bibinfo {year} {1988})%
  \bibAnnoteFile{NoStop}{Wees1988}%
\bibitem{Topinka2000}%
  \BibitemOpen
  \bibfield{author}{%
  \bibinfo {author} {\bibfnamefont{M.~A.}\ \bibnamefont{Topinka}}, \bibinfo
  {author} {\bibfnamefont{B.~J.}\ \bibnamefont{LeRoy}}, \bibinfo {author}
  {\bibfnamefont{S.~E.~J.}\ \bibnamefont{Shaw}}, \bibinfo {author}
  {\bibfnamefont{E.~J.}\ \bibnamefont{Heller}}, \bibinfo {author}
  {\bibfnamefont{R.~M.}\ \bibnamefont{Westervelt}}, \bibinfo {author}
  {\bibfnamefont{K.~D.}\ \bibnamefont{Maranowski}},\ and\ \bibinfo {author}
  {\bibfnamefont{A.~C.}\ \bibnamefont{Gossard}},\ }%
  \bibfield{journal}{%
  \Doi{10.1126/science.289.5488.2323}{\bibinfo {journal} {Science}}\ }%
  \textbf{\bibinfo {volume} {289}},\ \bibinfo {pages} {2323} (\bibinfo {year}
  {2000})%
  \bibAnnoteFile{NoStop}{Topinka2000}%
\bibitem{Topinka2001}%
  \BibitemOpen
  \bibfield{author}{%
  \bibinfo {author} {\bibfnamefont{M.~A.}\ \bibnamefont{Topinka}}, \bibinfo
  {author} {\bibfnamefont{B.~J.}\ \bibnamefont{LeRoy}}, \bibinfo {author}
  {\bibfnamefont{R.~M.}\ \bibnamefont{Westervelt}}, \bibinfo {author}
  {\bibfnamefont{S.~E.~J.}\ \bibnamefont{Shaw}}, \bibinfo {author}
  {\bibfnamefont{R.}~\bibnamefont{Fleischmann}}, \bibinfo {author}
  {\bibfnamefont{E.~J.}\ \bibnamefont{Heller}}, \bibinfo {author}
  {\bibfnamefont{K.~D.}\ \bibnamefont{Maranowski}},\ and\ \bibinfo {author}
  {\bibfnamefont{A.~C.}\ \bibnamefont{Gossard}},\ }%
  \bibfield{journal}{%
  \Doi{10.1038/35065553}{\bibinfo {journal} {Nature}}\ }%
  \textbf{\bibinfo {volume} {410}},\ \bibinfo {pages} {183} (\bibinfo {year}
  {2001})%
  \bibAnnoteFile{NoStop}{Topinka2001}%
\bibitem{Kozikov2015}%
  \BibitemOpen
  \bibfield{author}{%
  \bibinfo {author} {\bibfnamefont{A.~A.}\ \bibnamefont{Kozikov}}, \bibinfo
  {author} {\bibfnamefont{R.}~\bibnamefont{Steinacher}}, \bibinfo {author}
  {\bibfnamefont{C.}~\bibnamefont{R\"ossler}}, \bibinfo {author}
  {\bibfnamefont{T.}~\bibnamefont{Ihn}}, \bibinfo {author}
  {\bibfnamefont{K.}~\bibnamefont{Ensslin}}, \bibinfo {author}
  {\bibfnamefont{C.}~\bibnamefont{Reichl}},\ and\ \bibinfo {author}
  {\bibfnamefont{W.}~\bibnamefont{Wegscheider}},\ }%
  \bibfield{journal}{%
  \Doi{10.1021/acs.nanolett.5b03170}{\bibinfo {journal} {Nano Lett.}}\ }%
  \textbf{\bibinfo {volume} {15}},\ \bibinfo {pages} {7994} (\bibinfo {year}
  {2015})%
  \bibAnnoteFile{NoStop}{Kozikov2015}%
\bibitem{Jura2007}%
  \BibitemOpen
  \bibfield{author}{%
  \bibinfo {author} {\bibfnamefont{M.~P.}\ \bibnamefont{Jura}}, \bibinfo
  {author} {\bibfnamefont{M.~A.}\ \bibnamefont{Topinka}}, \bibinfo {author}
  {\bibfnamefont{L.}~\bibnamefont{Urban}}, \bibinfo {author}
  {\bibfnamefont{A.}~\bibnamefont{Yazdani}}, \bibinfo {author}
  {\bibfnamefont{H.}~\bibnamefont{Shtrikman}}, \bibinfo {author}
  {\bibfnamefont{L.~N.}\ \bibnamefont{Pfeiffer}}, \bibinfo {author}
  {\bibfnamefont{K.~W.}\ \bibnamefont{West}},\ and\ \bibinfo {author}
  {\bibfnamefont{D.}~\bibnamefont{Goldhaber-Gordon}},\ }%
  \bibfield{journal}{%
  \Doi{10.1038/nphys756}{\bibinfo {journal} {Nat Phys}}\ }%
  \textbf{\bibinfo {volume} {3}},\ \bibinfo {pages} {841} (\bibinfo {year}
  {2007}),\ ISSN \bibinfo {issn} {1745-2473}%
  \bibAnnoteFile{NoStop}{Jura2007}%
\bibitem{Jura2009}%
  \BibitemOpen
  \bibfield{author}{%
  \bibinfo {author} {\bibfnamefont{M.~P.}\ \bibnamefont{Jura}}, \bibinfo
  {author} {\bibfnamefont{M.~A.}\ \bibnamefont{Topinka}}, \bibinfo {author}
  {\bibfnamefont{M.}~\bibnamefont{Grobis}}, \bibinfo {author}
  {\bibfnamefont{L.~N.}\ \bibnamefont{Pfeiffer}}, \bibinfo {author}
  {\bibfnamefont{K.~W.}\ \bibnamefont{West}},\ and\ \bibinfo {author}
  {\bibfnamefont{D.}~\bibnamefont{Goldhaber-Gordon}},\ }%
  \bibfield{journal}{%
  \Doi{10.1103/PhysRevB.80.041303}{\bibinfo {journal} {Phys. Rev. B}}\ }%
  \textbf{\bibinfo {volume} {80}},\ \bibinfo {pages} {041303} (\bibinfo {year}
  {2009})%
  \bibAnnoteFile{NoStop}{Jura2009}%
\bibitem{Kumar2013}%
  \BibitemOpen
  \bibfield{author}{%
  \bibinfo {author} {\bibfnamefont{M.}~\bibnamefont{Kumar}}, \bibinfo {author}
  {\bibfnamefont{S.}~\bibnamefont{Lahon}}, \bibinfo {author}
  {\bibfnamefont{P.~K.}\ \bibnamefont{Jha}},\ and\ \bibinfo {author}
  {\bibfnamefont{M.}~\bibnamefont{Mohan}},\ }%
  \bibfield{journal}{%
  \Doi{http://dx.doi.org/10.1016/j.spmi.2013.01.007}{\bibinfo {journal}
  {Superlattices Microstruct.}}\ }%
  \textbf{\bibinfo {volume} {57}},\ \bibinfo {pages} {11 } (\bibinfo {year}
  {2013})%
  \bibAnnoteFile{NoStop}{Kumar2013}%
\bibitem{Heller2003}%
  \BibitemOpen
  \bibfield{author}{%
  \bibinfo {author} {\bibfnamefont{E.~J.}\ \bibnamefont{Heller}}\ and\ \bibinfo
  {author} {\bibfnamefont{S.}~\bibnamefont{Shaw}},\ }%
  \bibfield{journal}{%
  \Doi{10.1142/S0217979203021964}{\bibinfo {journal} {Int. J. Mod. Phys. B}}\
  }%
  \textbf{\bibinfo {volume} {17}},\ \bibinfo {pages} {3977} (\bibinfo {year}
  {2003})%
  \bibAnnoteFile{NoStop}{Heller2003}%
\bibitem{Liu2013}%
  \BibitemOpen
  \bibfield{author}{%
  \bibinfo {author} {\bibfnamefont{B.}~\bibnamefont{Liu}}\ and\ \bibinfo
  {author} {\bibfnamefont{E.~J.}\ \bibnamefont{Heller}},\ }%
  \bibfield{journal}{%
  \Doi{10.1103/PhysRevLett.111.236804}{\bibinfo {journal} {Phys. Rev. Lett.}}\
  }%
  \textbf{\bibinfo {volume} {111}},\ \bibinfo {pages} {236804} (\bibinfo
  {month} {Dec}\ \bibinfo {year} {2013})%
  \bibAnnoteFile{NoStop}{Liu2013}%
\bibitem{Liu2015}%
  \BibitemOpen
  \bibfield{author}{%
  \bibinfo {author} {\bibfnamefont{B.}~\bibnamefont{Liu}},\ }%
  \bibfield{journal}{%
  \bibinfo {journal} {Journal of Physics: Conference Series}\ }%
  \textbf{\bibinfo {volume} {626}},\ \bibinfo {pages} {012037} (\bibinfo {year}
  {2015})%
  \bibAnnoteFile{NoStop}{Liu2015}%
\bibitem{Kolasinski2014Slit}%
  \BibitemOpen
  \bibfield{author}{%
  \bibinfo {author}
  {\bibfnamefont{K.}~\bibnamefont{Kolasi\ifmmode~\acute{n}\else
  \'{n}\fi{}ski}}, \bibinfo {author}
  {\bibfnamefont{B.}~\bibnamefont{Szafran}},\ and\ \bibinfo {author}
  {\bibfnamefont{M.~P.}\ \bibnamefont{Nowak}},\ }%
  \bibfield{journal}{%
  \Doi{10.1103/PhysRevB.90.165303}{\bibinfo {journal} {Phys. Rev. B}}\ }%
  \textbf{\bibinfo {volume} {90}},\ \bibinfo {pages} {165303} (\bibinfo {year}
  {2014})%
  \bibAnnoteFile{NoStop}{Kolasinski2014Slit}%
\bibitem{Davies1995}%
  \BibitemOpen
  \bibfield{author}{%
  \bibinfo {author} {\bibfnamefont{J.~H.}\ \bibnamefont{Davies}}, \bibinfo
  {author} {\bibfnamefont{I.~A.}\ \bibnamefont{Larkin}},\ and\ \bibinfo
  {author} {\bibfnamefont{E.~V.}\ \bibnamefont{Sukhorukov}},\ }%
  \bibfield{journal}{%
  \bibinfo {journal} {J. Appl. Phys}\ }%
  \textbf{\bibinfo {volume} {77}},\ \bibinfo {pages} {4504} (\bibinfo {year}
  {1995})%
  \bibAnnoteFile{NoStop}{Davies1995}%
\bibitem{kolasinskiDFT2013}%
  \BibitemOpen
  \bibfield{author}{%
  \bibinfo {author}
  {\bibfnamefont{K.}~\bibnamefont{Kolasi\ifmmode~\acute{n}\else
  \'{n}\fi{}ski}}\ and\ \bibinfo {author}
  {\bibfnamefont{B.}~\bibnamefont{Szafran}},\ }%
  \bibfield{journal}{%
  \Doi{10.1103/PhysRevB.88.165306}{\bibinfo {journal} {Phys. Rev. B}}\ }%
  \textbf{\bibinfo {volume} {88}},\ \bibinfo {pages} {165306} (\bibinfo {year}
  {2013})%
  \bibAnnoteFile{NoStop}{kolasinskiDFT2013}%
\bibitem{szafranDFT2011}%
  \BibitemOpen
  \bibfield{author}{%
  \bibinfo {author} {\bibfnamefont{B.}~\bibnamefont{Szafran}},\ }%
  \bibfield{journal}{%
  \Doi{10.1103/PhysRevB.84.075336}{\bibinfo {journal} {Phys. Rev. B}}\ }%
  \textbf{\bibinfo {volume} {84}},\ \bibinfo {pages} {075336} (\bibinfo {year}
  {2011})%
  \bibAnnoteFile{NoStop}{szafranDFT2011}%
\bibitem{Steinacher2015}%
  \BibitemOpen
  \bibfield{author}{%
  \bibinfo {author} {\bibfnamefont{R.}~\bibnamefont{Steinacher}}, \bibinfo
  {author} {\bibfnamefont{A.~A.}\ \bibnamefont{Kozikov}}, \bibinfo {author}
  {\bibfnamefont{C.}~\bibnamefont{R\"ossler}}, \bibinfo {author}
  {\bibfnamefont{C.}~\bibnamefont{Reichl}}, \bibinfo {author}
  {\bibfnamefont{W.}~\bibnamefont{Wegscheider}}, \bibinfo {author}
  {\bibfnamefont{T.}~\bibnamefont{Ihn}},\ and\ \bibinfo {author}
  {\bibfnamefont{K.}~\bibnamefont{Ensslin}},\ }%
  \bibfield{journal}{%
  \bibinfo {journal} {New J. Phys.}\ }%
  \textbf{\bibinfo {volume} {17}},\ \bibinfo {pages} {043043} (\bibinfo {year}
  {2015})%
  \bibAnnoteFile{NoStop}{Steinacher2015}%
\bibitem{Zwierzycki2008}%
  \BibitemOpen
  \bibfield{author}{%
  \bibinfo {author} {\bibfnamefont{M.}~\bibnamefont{Zwierzycki}}, \bibinfo
  {author} {\bibfnamefont{P.~A.}\ \bibnamefont{Khomyakov}}, \bibinfo {author}
  {\bibfnamefont{A.~A.}\ \bibnamefont{Starikov}}, \bibinfo {author}
  {\bibfnamefont{K.}~\bibnamefont{Xia}}, \bibinfo {author}
  {\bibfnamefont{M.}~\bibnamefont{Talanana}}, \bibinfo {author}
  {\bibfnamefont{P.~X.}\ \bibnamefont{Xu}}, \bibinfo {author}
  {\bibfnamefont{V.~M.}\ \bibnamefont{Karpan}}, \bibinfo {author}
  {\bibfnamefont{I.}~\bibnamefont{Marushchenko}}, \bibinfo {author}
  {\bibfnamefont{I.}~\bibnamefont{Turek}}, \bibinfo {author}
  {\bibfnamefont{G.~E.~W.}\ \bibnamefont{Bauer}}, \bibinfo {author}
  {\bibfnamefont{G.}~\bibnamefont{Brocks}},\ and\ \bibinfo {author}
  {\bibfnamefont{P.~J.}\ \bibnamefont{Kelly}},\ }%
  \bibfield{journal}{%
  \Doi{10.1002/pssb.200743359}{\bibinfo {journal} {Phys. Stat. Sol.}}\ }%
  \textbf{\bibinfo {volume} {245}},\ \bibinfo {pages} {623} (\bibinfo {year}
  {2008})%
  \bibAnnoteFile{NoStop}{Zwierzycki2008}%
\bibitem{AndoWFM1991}%
  \BibitemOpen
  \bibfield{author}{%
  \bibinfo {author} {\bibfnamefont{T.}~\bibnamefont{Ando}},\ }%
  \bibfield{journal}{%
  \Doi{10.1103/PhysRevB.44.8017}{\bibinfo {journal} {Phys. Rev. B}}\ }%
  \textbf{\bibinfo {volume} {44}},\ \bibinfo {pages} {8017} (\bibinfo {year}
  {1991})%
  \bibAnnoteFile{NoStop}{AndoWFM1991}%
\bibitem{Khomaykov2005}%
  \BibitemOpen
  \bibfield{author}{%
  \bibinfo {author} {\bibfnamefont{P.~A.}\ \bibnamefont{Khomyakov}}, \bibinfo
  {author} {\bibfnamefont{G.}~\bibnamefont{Brocks}}, \bibinfo {author}
  {\bibfnamefont{V.}~\bibnamefont{Karpan}}, \bibinfo {author}
  {\bibfnamefont{M.}~\bibnamefont{Zwierzycki}},\ and\ \bibinfo {author}
  {\bibfnamefont{P.~J.}\ \bibnamefont{Kelly}},\ }%
  \bibfield{journal}{%
  \Doi{10.1103/PhysRevB.72.035450}{\bibinfo {journal} {Phys. Rev. B}}\ }%
  \textbf{\bibinfo {volume} {72}},\ \bibinfo {pages} {035450} (\bibinfo {year}
  {2005})%
  \bibAnnoteFile{NoStop}{Khomaykov2005}%
\bibitem{Paradiso2010}%
  \BibitemOpen
  \bibfield{author}{%
  \bibinfo {author} {\bibfnamefont{N.}~\bibnamefont{Paradiso}}, \bibinfo
  {author} {\bibfnamefont{S.}~\bibnamefont{Heun}}, \bibinfo {author}
  {\bibfnamefont{S.}~\bibnamefont{Roddaro}}, \bibinfo {author}
  {\bibfnamefont{L.}~\bibnamefont{Pfeiffer}}, \bibinfo {author}
  {\bibfnamefont{K.}~\bibnamefont{West}}, \bibinfo {author}
  {\bibfnamefont{L.}~\bibnamefont{Sorba}}, \bibinfo {author}
  {\bibfnamefont{G.}~\bibnamefont{Biasiol}},\ and\ \bibinfo {author}
  {\bibfnamefont{F.}~\bibnamefont{Beltram}},\ }%
  \bibfield{journal}{%
  \Doi{http://dx.doi.org/10.1016/j.physe.2009.11.146}{\bibinfo {journal}
  {Physica E}}\ }%
  \textbf{\bibinfo {volume} {42}},\ \bibinfo {pages} {1038 } (\bibinfo {year}
  {2010}),\ \bibinfo {note} {18th International Conference on Electron
  Properties of Two-Dimensional Systems}%
  \bibAnnoteFile{NoStop}{Paradiso2010}%
\bibitem{Kozikov2013}%
  \BibitemOpen
  \bibfield{author}{%
  \bibinfo {author} {\bibfnamefont{A.~A.}\ \bibnamefont{Kozikov}}, \bibinfo
  {author} {\bibfnamefont{C.}~\bibnamefont{Rössler}}, \bibinfo {author}
  {\bibfnamefont{T.}~\bibnamefont{Ihn}}, \bibinfo {author}
  {\bibfnamefont{K.}~\bibnamefont{Ensslin}}, \bibinfo {author}
  {\bibfnamefont{C.}~\bibnamefont{Reichl}},\ and\ \bibinfo {author}
  {\bibfnamefont{W.}~\bibnamefont{Wegscheider}},\ }%
  \bibfield{journal}{%
  \bibinfo {journal} {New Journal of Physics}\ }%
  \textbf{\bibinfo {volume} {15}},\ \bibinfo {pages} {013056} (\bibinfo {year}
  {2013})%
  \bibAnnoteFile{NoStop}{Kozikov2013}%
\bibitem{Steinacher2016}%
  \BibitemOpen
  \bibfield{author}{%
  \bibinfo {author} {\bibfnamefont{R.}~\bibnamefont{Steinacher}}, \bibinfo
  {author} {\bibfnamefont{A.~A.}\ \bibnamefont{Kozikov}}, \bibinfo {author}
  {\bibfnamefont{C.}~\bibnamefont{R\"ossler}}, \bibinfo {author}
  {\bibfnamefont{C.}~\bibnamefont{Reichl}}, \bibinfo {author}
  {\bibfnamefont{W.}~\bibnamefont{Wegscheider}}, \bibinfo {author}
  {\bibfnamefont{K.}~\bibnamefont{Ensslin}},\ and\ \bibinfo {author}
  {\bibfnamefont{T.}~\bibnamefont{Ihn}},\ }%
  \bibfield{journal}{%
  \Doi{10.1103/PhysRevB.93.085303}{\bibinfo {journal} {Phys. Rev. B}}\ }%
  \textbf{\bibinfo {volume} {93}},\ \bibinfo {pages} {085303} (\bibinfo {year}
  {2016})%
  \bibAnnoteFile{NoStop}{Steinacher2016}%
\bibitem{Brun2016}%
  \BibitemOpen
  \bibfield{author}{%
  \bibinfo {author} {\bibfnamefont{B.}~\bibnamefont{Brun}}, \bibinfo {author}
  {\bibfnamefont{F.}~\bibnamefont{Martins}}, \bibinfo {author}
  {\bibfnamefont{S.}~\bibnamefont{Faniel}}, \bibinfo {author}
  {\bibfnamefont{B.}~\bibnamefont{Hackens}}, \bibinfo {author}
  {\bibfnamefont{A.}~\bibnamefont{Cavanna}}, \bibinfo {author}
  {\bibfnamefont{C.}~\bibnamefont{Ulysse}}, \bibinfo {author}
  {\bibfnamefont{A.}~\bibnamefont{Ouerghi}}, \bibinfo {author}
  {\bibfnamefont{U.}~\bibnamefont{Gennser}}, \bibinfo {author}
  {\bibfnamefont{D.}~\bibnamefont{Mailly}}, \bibinfo {author}
  {\bibfnamefont{P.}~\bibnamefont{Simon}}, \bibinfo {author}
  {\bibfnamefont{S.}~\bibnamefont{Huant}}, \bibinfo {author}
  {\bibfnamefont{V.}~\bibnamefont{Bayot}}, \bibinfo {author}
  {\bibfnamefont{M.}~\bibnamefont{Sanquer}},\ and\ \bibinfo {author}
  {\bibfnamefont{H.}~\bibnamefont{Sellier}},\ }%
  \bibfield{journal}{%
  \bibinfo {journal} {Phys. Rev. Lett}}%
   (\bibinfo {year} {2016}),\ \bibinfo {note} {(to be published)}%
  \bibAnnoteFile{NoStop}{Brun2016}%
\bibitem{Kirkner1990}%
  \BibitemOpen
  \bibfield{author}{%
  \bibinfo {author} {\bibfnamefont{C.~S.}\ \bibnamefont{Lent}}\ and\ \bibinfo
  {author} {\bibfnamefont{D.~J.}\ \bibnamefont{Kirkner}},\ }%
  \bibfield{journal}{%
  \Doi{http://dx.doi.org/10.1063/1.345156}{\bibinfo {journal} {J. Appl.
  Phys.}}\ }%
  \textbf{\bibinfo {volume} {67}},\ \bibinfo {pages} {6353} (\bibinfo {year}
  {1990})%
  \bibAnnoteFile{NoStop}{Kirkner1990}%
\bibitem{Kolasinski2016Lande}%
  \BibitemOpen
  \bibfield{author}{%
  \bibinfo {author}
  {\bibfnamefont{K.}~\bibnamefont{Kolasi\ifmmode~\acute{n}\else
  \'{n}\fi{}ski}}, \bibinfo {author}
  {\bibfnamefont{A.}~\bibnamefont{Mre\ifmmode \acute{n}\else
  \'{n}\fi{}ca-Kolasi\ifmmode~\acute{n}\else \'{n}\fi{}ska}},\ and\ \bibinfo
  {author} {\bibfnamefont{B.}~\bibnamefont{Szafran}},\ }%
  \bibfield{journal}{%
  \Doi{10.1103/PhysRevB.93.035304}{\bibinfo {journal} {Phys. Rev. B}}\ }%
  \textbf{\bibinfo {volume} {93}},\ \bibinfo {pages} {035304} (\bibinfo {year}
  {2016})%
  \bibAnnoteFile{NoStop}{Kolasinski2016Lande}%
\bibitem{Rungger2008}%
  \BibitemOpen
  \bibfield{author}{%
  \bibinfo {author} {\bibfnamefont{I.}~\bibnamefont{Rungger}}\ and\ \bibinfo
  {author} {\bibfnamefont{S.}~\bibnamefont{Sanvito}},\ }%
  \bibfield{journal}{%
  \Doi{10.1103/PhysRevB.78.035407}{\bibinfo {journal} {Phys. Rev. B}}\ }%
  \textbf{\bibinfo {volume} {78}},\ \bibinfo {pages} {035} (\bibinfo {year}
  {2008})%
  \bibAnnoteFile{NoStop}{Rungger2008}%
\end{thebibliography}%

\end{document}